# StimDust: A mm-scale implantable wireless precision neural stimulator with ultrasonic power and communication

David K. Piech[1]*, Benjamin C. Johnson[2,3]*, Konlin Shen[1], M. Meraj Ghanbari[2], Ka Yiu Li[2], Ryan M. Neely[4], Joshua E. Kay[2], Jose M. Carmena[2,4]**, Michel M. Maharbiz[2,5]**, Rikky Muller[2,5]**

[1] University of California, Berkeley – University of California, San Francisco Graduate Program in Bioengineering, University of California, Berkeley, Berkeley, CA USA 94720

[2] Department of Electrical Engineering and Computer Sciences, University of California, Berkeley, Berkeley, CA USA 94720

[3] Department of Electrical Engineering and Computer Engineering, Boise State University, Boise, ID USA 83725

[4] Helen Wills Neuroscience Institute, University of California, Berkeley, Berkeley, CA USA 94720

[5] Chan-Zuckerberg Biohub, San Francisco, CA USA 94158

* Co-first authors: D.K.P., B.C.J.;    ** Co-senior authors: R.M., M.M.M., J.M.C.

** e-mail: R.M. (rikky@berkeley.edu); M.M.M. (maharbiz@berkeley.edu); J.M.C. (jcarmena@berkeley.edu)

## Abstract

Neural stimulation is a powerful technique for modulating physiological functions and for writing information into the nervous system as part of brain-machine interfaces. Current clinically approved neural stimulators require batteries and are many cubic centimetres in size- typically much larger than their intended targets. We present a complete wireless neural stimulation system consisting of a 1.7 mm$^3$ wireless, batteryless, leadless implantable stimulator (the "mote"), an ultrasonic wireless link for power and bi-directional communication, and a hand-held external transceiver. The mote consists of a piezoceramic transducer, an energy storage capacitor, and a stimulator integrated circuit (IC). The IC harvests ultrasonic power with high efficiency, decodes stimulation parameter downlink data, and generates current-controlled stimulation pulses. Stimulation parameters are time-encoded on the fly through the wireless link rather than being programmed and stored on the mote, reducing power consumption and on-chip memory requirements and enabling complex stimulation protocols with high-temporal resolution and low-latency feedback for use in closed-loop stimulation. Uplink data indicates whether the mote is currently stimulating; it is encoded by the mote via backscatter modulation and is demodulated at the external transceiver. We show that the mote operates at an acoustic intensity that is 7.8% of the FDA limit for diagnostic ultrasound and characterize the acoustic wireless link's robustness to expected real-world misalignment. We demonstrate the *in vivo* performance of the system with motes acutely implanted with a cuff on the sciatic nerve of anesthetized rats and show highly repeatable stimulation across a wide range of physiological responses.

Extracellular electrophysiological stimulation is broadly applied to targets in the peripheral and central nervous system and has a long history of application both in neuroscience and clinical therapies. In the peripheral nervous system (PNS), the cochlear implant is well established for direct stimulation of auditory sensory afferents and the creation of useful auditory percepts[1]. More recent work has focused on creating devices that directly stimulate the PNS to regulate physiological functions[2] such as modulating blood pressure[3], rheumatoid arthritis[4], incontinence[5], sexual function[6], and immune system function[7]; and directly writing neural codes to sensory afferents[8] and motor efferents[9]. Electrical stimulation of central nervous system (CNS) targets is well-established for treatment of chronic pain[10] and central motor disorders such as tremor in Parkinson's disease[11] with deep-brain stimulation (DBS). Recent work shows promising results for treatment of depression[12], modulation of decision making[13], and use in closed-loop brain-machine interfacing[14,15].

New applications, use-cases, and patient populations for neural stimulation would be opened up by stimulators which have significantly reduced total implant size, lessened implantation risk, improved tissue interface longevity, and no transcutaneous wires. Current clinically deployed neural stimulators have a volume greater than 6000 mm$^3$ [16,17,18] and consequently must be placed in large anatomical pockets. These pockets can be far from the stimulation site (e.g. a subclavical stimulator with a lead to the subthalamic nucleus) or require removing substantial soft tissue or bone to form a pocket (e.g. Neuropace RNS[18]). While studies have shown that clinically-relevant stimulation performance can be achieved for several years with existing technologies[19, 20, 8, 21], technologies which have potential for further improving longevity would be advantageous. Long leads are required from the stimulator housing to the stimulation site, which is a source of infection, efficacy loss, and device failure[22], in addition to electromagnetic interference, crosstalk, and power inefficiencies. Peripheral nerve cuff electrodes which use wires to connect to a distant stimulator implant have shown fatigue and breakage in the lead[23], emphasizing the benefit of a completely untethered device. A stimulator of a couple cubic millimetres in size, comparable to a grain of sand, could be entirely and independently placed directly at the site of stimulation rather than requiring leads. In comparison to traditional devices which require greater anatomical space, tunnelling for a lead and/or an anatomical pocked for a relatively large stimulator package, a highly miniaturized wireless stimulator may improve safety, access to anatomical sites, and enable ultra-minimally-invasive delivery methods such as laparoscopy or injection, reducing tissue trauma during implantation and immune response[24] Such a device may improve the reward-to-risk ratio of placing stimulators in certain anatomical locations, opening up new treatments and patient populations.

A key challenge in developing wireless mm-scale stimulators is wireless power and communication that provides well-controlled, therapeutically relevant effects. Cochlear implant stimulators, have demonstrated wireless neural stimulation for decades[25], but they are large (6000 mm$^3$) and operate only at shallow depth due to the use of an inductive link[26,27]. To date, the smallest volume stimulators have been implemented with passive components, providing inefficient voltage-mode stimulation that is sensitive to received power and electrode impedance[28,29]. Additionally, these stimulators do not have an uplink, preventing error detection or state monitoring via feedback to the external controller. Precise control of stimulation charge is important for targeting a repeatable population of cells and maintaining safety over chronic use. Neurostimulators employing integrated circuits can use active rectifiers to achieve a higher power conversion efficiency and deliver current stimulation that is independent of received power[30,31,32]. However, most active neurostimulators rely on a recovered clock from the high frequency wireless signal as well as programmed registers to generate stimulation waveform timing. Significant division of the recovered clock results in power overhead and programming large registers limits the dynamic programmability of the waveform. Furthermore, most of these wireless systems use electromagnetic (EM) waves for power delivery and communication, which couples inefficiently at centimetre scales deep within tissue [33].

Recently, ultrasound has been proposed as an efficient way to power and communicate with mm-scale implants deep in tissue[34], resulting in volume reduction for implants[35,36,37,38] which maintain the ability to operate at moderate depths in tissue and perform functions with moderate power requirements. Compared to electromagnetic energy at 2 GHz, ultrasound energy at 2 MHz has less attenuation through tissue (1-2 dB/cm vs. 10-12 dB/cm[39]), a higher FDA (United States Food and Drug Administration) limit for power flux (7.2 mW/mm$^2$ vs. 0.1 mW/mm$^2$ [40]), and a smaller wavelength (0.75 mm vs. 25 mm) for more efficient coupling to small implants.

Here, we present a system for wireless neural stimulation that employs 1.7 mm$^3$ leadless, wireless "StimDust" stimulation motes. The ultrasonically powered StimDust motes contain a state-of-the-art custom integrated circuit (IC). A single link from a custom external transceiver to the mote provides power and bidirectional communication, reducing volume. A custom, time-coded wireless ultrasonic protocol dynamically sets stimulation waveforms, drastically reducing mote power consumption while maintaining a high degree of timing control.

# Results

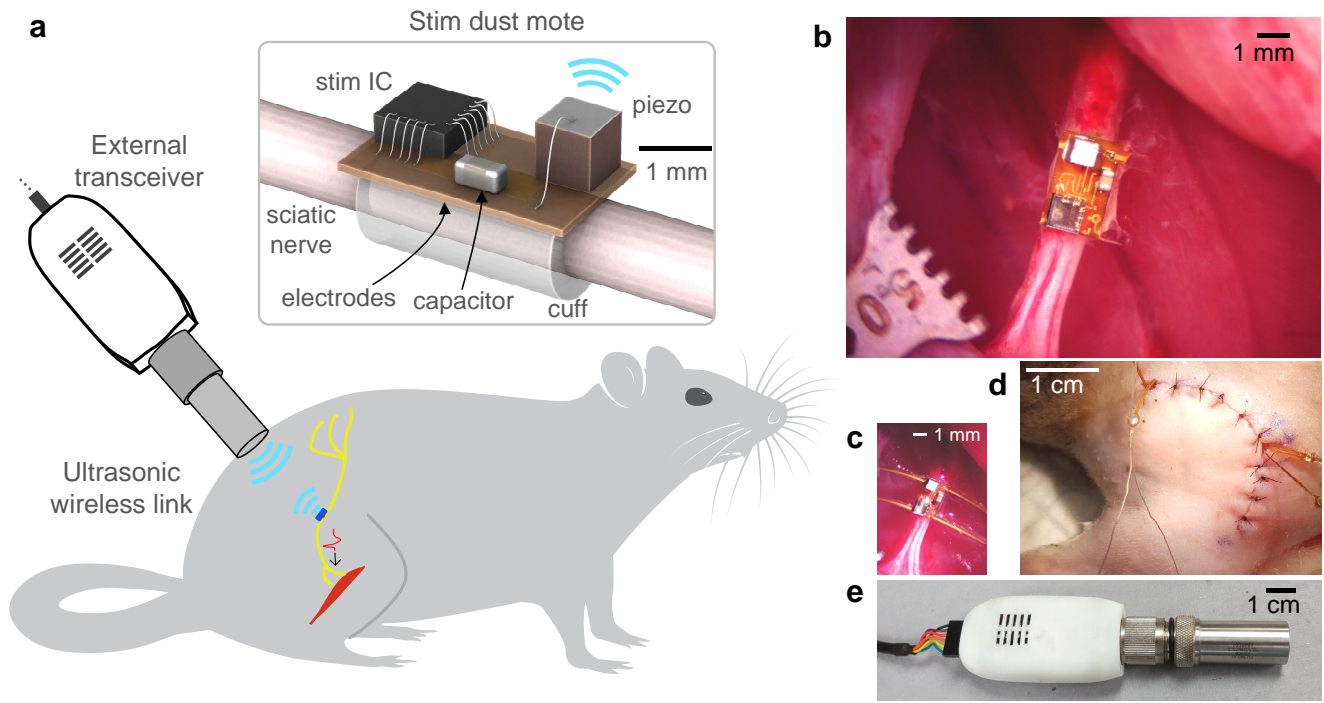

**Fig. 1 | StimDust wireless neural stimulator system overview. a**, Diagram depicting the StimDust system; here, it is used for stimulation of the sciatic nerve of a rat. **b**, The stimulating mote fully implanted and affixed to a rat sciatic nerve. Piezo at top, IC at bottom of mote. Stimulation pulses from the mote elicited compound action potentials in the nerve and compound muscle action potentials in the hindlimb musculature. (One of 58 images taken; see Fig S11) **c**, Optional test leads on the mote were used for diagnostic and data reporting purposes to measure internal signals on the mote (also seen in (d)). They provided no power or control signals to the mote, and the mote function was verified with these leads disconnected from all external instrumentation. (One of 12 images taken; see Fig S11). **d**, The wireless link traversed skin and muscle on top of a closed surgical site; this image shows the hindquarters of the animal superficial to the sciatic nerve. (One of 49 images taken). **e**, The external transceiver established an ultrasonic wireless link with the mote. The power and data cable is shown on the left and the external transducer is on the right. (Representative image of 16 taken). **b** and **c** from animal A; **d** from animal D.

A conceptual overview of the StimDust system and its implementation is shown in Fig. 1. An external transceiver (Fig. 1e) established an ultrasonic wireless link through body tissue to an implanted StimDust mote (Fig. 1b, c). A single ultrasonic link provided both power to the mote and bidirectional communication. The mote consisted of a piezoceramic as the ultrasonic transducer; a custom IC ("Stim IC") for power rectification, communication, and stimulation; a discrete capacitor for energy storage; bipolar stimulation electrodes; and a nerve cuff integrated on a thin polyimide printed circuit board (PCB). The mote was surgically implanted directly at the stimulation site, in this case the rat sciatic nerve (Fig. 1b), the wound was closed (Fig. 1d); the transcutaneous wires are optional test leads), and the external transceiver was coupled to the outside surface of the body with ultrasound gel. The system could produce a broad set of clinically relevant stimulation currents, pulse widths, and pulse repetition frequencies (PRF); these were sufficient to elicit highly repeatable compound action potentials (CAPs) in the sciatic nerve, compound muscle action potentials (CMAPs), and associated twitches in downstream muscles.

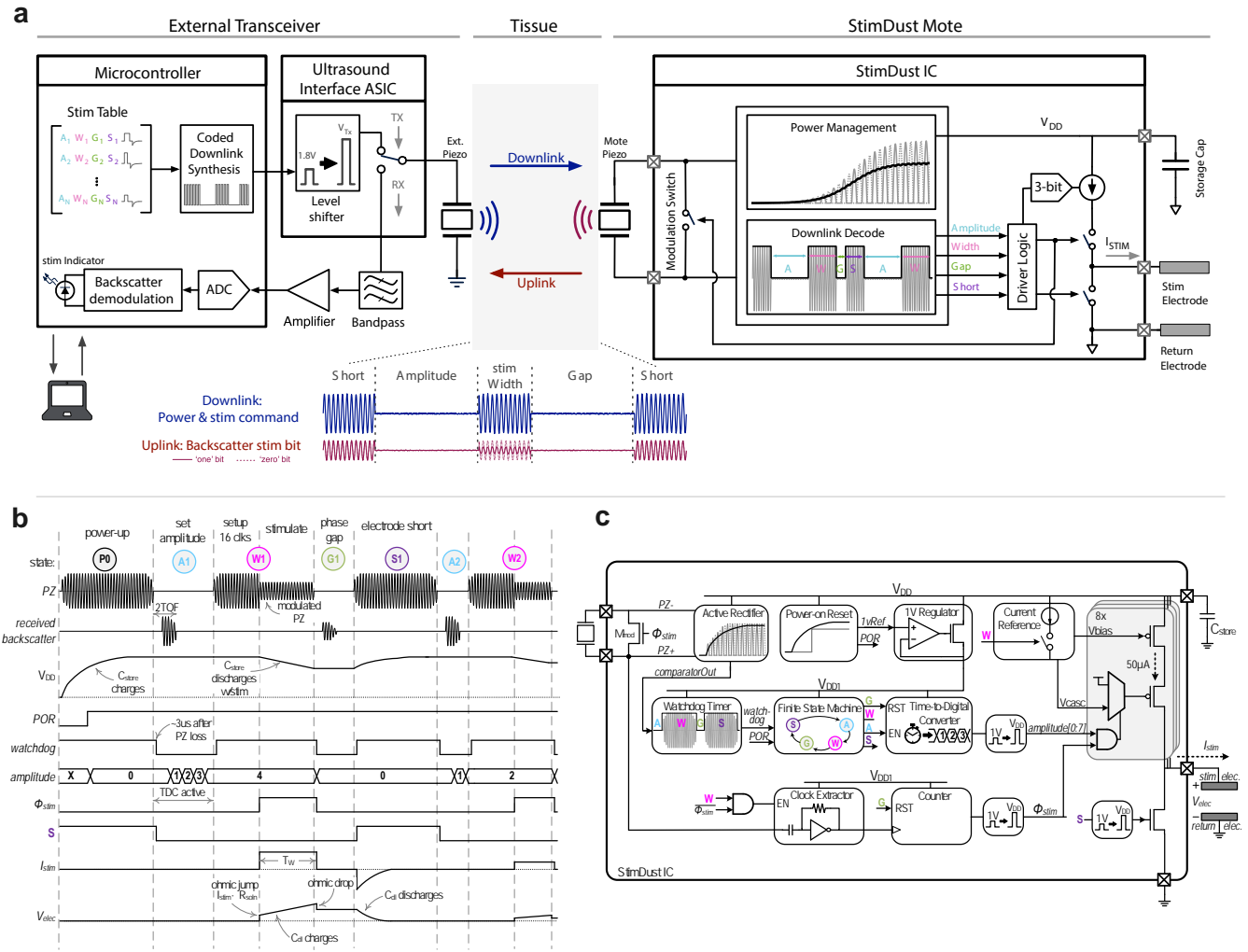

**Fig. 2 | StimDust system block diagram. a,** System overview: The external transceiver is depicted on the left, the ultrasonic wireless link in the middle, and the StimDust mote on the right. **left**, The external transceiver consisted of a transmit (Tx) path (left-top) which encoded user-specified stimulation parameters onto a 1.85 MHz carrier wave, stepped up the voltage, and drove an external piezo transducer during the Tx sequence. A receive (Rx) path (left-middle) was activated when the Tx was not active and captured and demodulated the ultrasonic backscatter. **middle**, The ultrasonic wireless link traversed the body tissue between the external transceiver and the implanted mote. The downlink provided power and time-delay coded stimulation commands for stim amplitude [A], stim pulse width [W], interphase gap duration [G], and electrode shorting duration [S]. The uplink consisted of backscatter amplitude modulation. **right**, The implanted mote harvested power and decoded stim parameters from the downlink. A finite state machine set stimulation parameters, initiated the specified stim pulse, and activated a modulation switch on the piezo input to change its ultrasound reflectance and encode a backscatter uplink bit indicating stim state information. **b**, The Stim IC timing diagram and **c**, architecture. Ultrasound from the external transducer charged VDD and the POR initialized the system. The watchdog detected ultrasound-free intervals and incremented the state. The amplitude was set by the duration of state A via a time-to-digital converter. $\Phi_{stim}$ enabled the stimulation current, $I_{stim}$, and the modulation switch, $M_{mod}$. Backscattered ultrasound was detected by the external transducer during the ultrasound-free intervals. IC circuitry was heavily duty-cycled to conserve power.

A functional block diagram of the system is detailed in Fig. 2. The external transceiver[41] (Fig. 1e, Fig 3g) utilized a microcontroller to synthesize a low-voltage transmit signal, which encoded downlink stimulation parameters for each pulse onto a 1.85 MHz carrier (1.85 MHz nominal; $f_{carrier}$ was tuned between 1.80 and 1.94 across the set of fabricated motes). The low-voltage transmit signal from the microcontroller was stepped-up in voltage and drove the external piezoelectric ultrasound transducer. During the 'off' periods of the transmit signal, the external transceiver toggled from Tx (transmit) to Rx (receive) mode to capture and demodulate uplink data from the acoustic signal backscattered by the mote.

The implanted mote used an IC[42] to rectify energy harvested by the mote's piezo. Rectified energy was stored on an off-chip capacitor to ensure consistent operation of the IC during the 'off' cycles. The IC decoded received ultrasound to determine stimulation current amplitude (50 µA to 400 µA with 3b resolution), pulse width (450 ns resolution), interphase gap (450 ns resolution), post-stimulation electrode shorting phase ('passive recharge', 450 ns resolution), and PRF (0 – 5 kHz with 450 ns resolution). The mote delivered current-mode stimulation, during which the IC modulated the impedance across the piezo to change the amplitude of the reflected wave. This indicated to the external transceiver when the IC was stimulating and operating properly.

**Protocol and implant design**

The Stim IC timing diagram and architecture are shown in Fig. 2b, c. The IC used a custom ultrasound protocol, which enabled dynamically controlled stimulation waveforms and high system efficiency while maintaining a high degree of timing control. The protocol utilized the shape of the incoming ultrasound envelope to encode stimulation parameters, eliminating the need for an on-chip, continuously running clock to ensure timing precision. This reduced power consumption and afforded greater stimulation waveform timing flexibility compared to solutions that rely on a clock receiver, divider circuits, and on-chip registers to generate the stimulation waveform[43,30].

As shown in Fig. 2, an initial power-up sequence charged the implant's storage capacitor, $C_{store}$. As the supply voltage of the IC ($V_{DD}$) ramped up to 2.5 V, a power-on-reset (POR) signal (threshold≈1.6 V) was generated to initialize the chip. If the POR signal was not generated, or $V_{DD}$ dropped below the threshold, the IC would not stimulate as a fail-safe against aberrant stimulation. In addition to the power-up sequence, the protocol was comprised of 4 repeating states: set amplitude (**A**), pulse width (**W**), interphase gap (**G**), and electrode shorting (**S**). Every state encoded information pertinent to the stimulation waveform interpreted by the IC and had a temporal resolution set by the carrier frequency

($1/f_{carrier}$ = 540 ns). States **A** and **G** are called ultrasound-free intervals (UFI). **A** occurred during the first ultrasound-free interval and its duration encoded the stimulation amplitude. The watchdog timer circuit detected the ultrasound-free interval, which then triggered a 3-bit time-to-digital converter (TDC). The 3-bits of the TDC output were stored in memory for the stimulation pulse amplitude. **W** began when ultrasound returned and started a 16-cycle setup sequence to allow the current reference to settle prior to stimulation. On the 16$^{th}$ cycle, a current mirror whose amplitude was set by the TDC was enabled by the output of the counter, $\Phi_{stim}$, causing the programmed current to flow from the stimulation electrode to the return electrode for a monophasic pulse.

Stimulation was terminated at the second ultrasound-free interval. During **G**, the chip was in standby mode and the stimulation electrode was tri-stated to create an interphase gap. Interphase gaps have been shown to decrease stimulation thresholds in auditory nerves, motor nerves, and retinal ganglion cells, in addition to decreasing perceptual thresholds in cochlear implant patients and epiretinal prosthesis patients[44,45,46,47,48,49]. State **S** started when ultrasound returned. During **S**, received ultrasound recharged the storage capacitor and the IC shorted the stimulation electrode to ground, clearing accumulated charge on the electrodes. Since the charge cleared quickly, it was possible to use monophasic stimulation with passive recharge (Supplementary Section 4, Fig S8a-g Fig S9, Fig S10, Table S2),[50] which consumes half the energy of conventional biphasic stimulation without limiting stimulation frequency in this work. Furthermore, biphasic stimulators typically use an additional shorting phase to clear residual charge that accumulates from phase mismatch[51] (See Supplementary Information section 4 for further detail).

Whenever ultrasound was on (**W** and **S**), a fraction of the incident ultrasonic energy was reflected to the external transceiver (backscatter). The backscatter amplitude was a function of the electrical load of the mote piezo. The delay between transceiver emission and backscatter detection was twice the acoustic time-of-flight (~20-60µs) between the external transducer and the mote piezo. During stimulation, a transistor switch, $M_{mod}$ (Fig. 2b), was turned on to create a weak short (~4 kΩ) across the piezo terminals, reducing the amplitude of the backscatter signal (Supplementary Fig. 1). When $M_{mod}$ was off, the electrical load impedance across the piezo ranged from 10 kΩ to 300 kΩ depending on the power consumption of the IC. The impedance change during stimulation provided a 1-bit data uplink and safety indicator, informing the external transducer when the mote was stimulating.

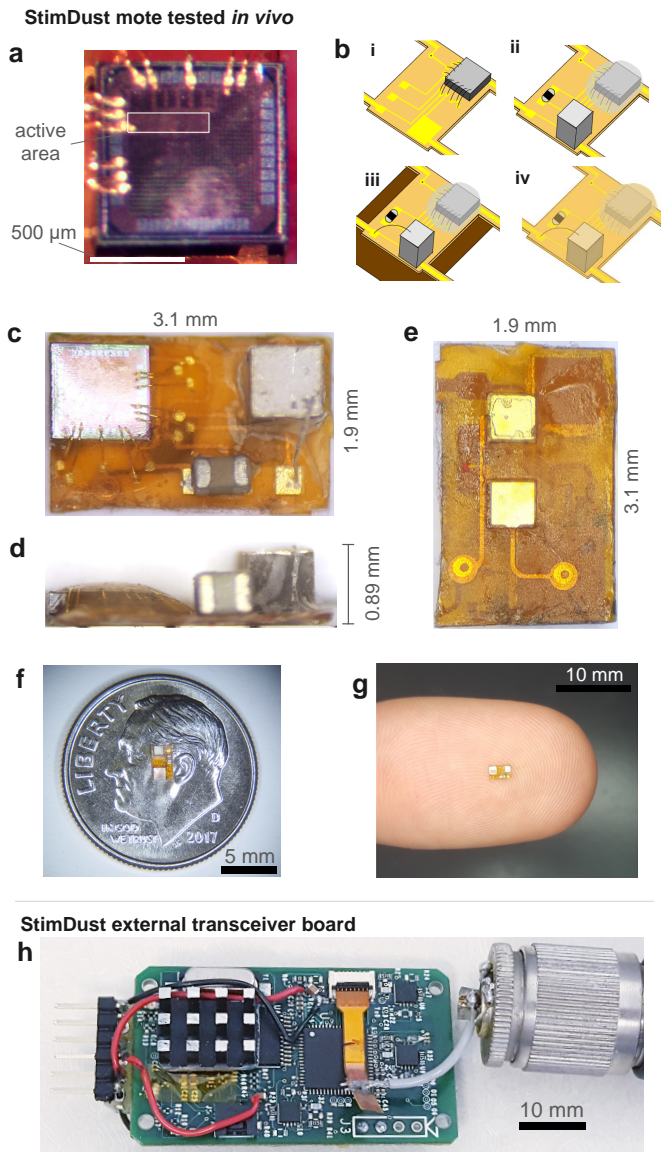

**Fig. 3 | StimDust fabrication. a**, Mote IC die photo (representative image of 9 taken). **b**, Mote fabrication steps: (i) IC die was wire-bonded to mote flex substrate; (ii) discrete capacitor and piezo were bonded with silver epoxy and die was encapsulated with epoxy; (iii) piezo was wire-bonded; (iv) mote was encapsulated with parylene. **c**, Top view of mote: piezo on left, capacitor at top, IC on right (representative image of 11 taken). **d**, Side view of mote: IC on left, capacitor and piezo on right (representative image of 33 taken). **e**, Underside view of mote: ground electrode on left, working electrode on right (representative image of 11 taken). **f**, Size comparison to US dime (representative image of 16 taken) and **g**, a finger (one of 36 images taken). **h**, External transceiver board (representative image of 4 taken). Data and power connection is on the left, microcontroller is under the passive heat sink, ultrasound interface chip is middle-right and transducer connector on the right.

StimDust motes (Fig. 3) consisted of the IC, external storage capacitor, and a 750 µm x 750 µm x 750 µm lead-zirconate titanate (PZT) piezoceramic transducer attached to a miniature flexible PCB and encapsulated with parylene. The piezo thickness was chosen such that the series resonance was at the carrier frequency of 1.85 MHz. This choice of thickness and carrier frequency was a trade-off between improving voltage harvest (larger thickness), lessening the effect of higher ultrasound attenuation at higher frequencies (lower carrier frequency), and reducing mote size (smaller thickness). The piezo thickness-to-width ratio (1:1) was a trade-off between improving current harvest (larger surface area), minimizing mote size (smaller surface area), and increasing ultrasound acceptance angle (smaller aspect ratio). Stimulation electrodes located on the bottom of the PCB were electroplated with poly(3,4-ethylenedioxythiophene) polystyrene sulfonate (PEDOT:PSS) to improve charge injection capacity and prevent compliance limiting when using the full current range of the chip.

The StimDust stimulator was $1.7 \pm 0.2$ mm$^3$ in volume. A cuff of $3.2 \pm 0.4$ mm$^3$ volume was added to affix the stimulator to the nerve for *in vivo* validation. A 5x higher current variant was also designed with the same footprint for applications with higher stimulation thresholds.

## Robust wireless operation *in vitro* and *ex vivo*

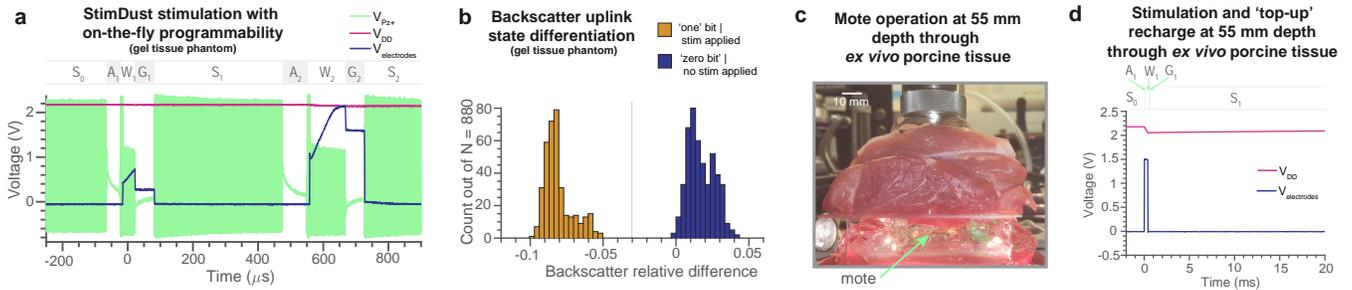

**Fig. 4 | StimDust demonstrated dynamic programmability, backscatter uplink communication, and operation at 55 mm depth through *ex-vivo* porcine tissue. a,** StimDust operation in a gel tissue phantom. Mote waveforms recorded during a representative stimulation sequence where stimulation parameters were changed on the fly, demonstrating dynamic programmability. The first stimulation pulse of 150 μA for 37 μs was specified by regions $A_1$, $W_1$, and $G_1$. The second stimulation pulse of 300 μA for 110 μs was specified by $A_2$, $W_2$, and $G_2$. These two different dynamically-encoded pulses are less than 500 μs apart. **b,** Backscatter relative difference (see Methods and Fig. 6e) for 878 stimulation pulses. A population of successful stim pulses with backscatter values of 'one' bits was taken under good mote operating conditions with verified stim at the outputs. A population of backscatter samples with no stim and backscatter values of 'zero' bits was acquired by operating the mote in an underpowered state such that output stimulation was no longer performed. The two populations show good separation, enabling the use of backscatter as an indicator of successful stimulation. These replications are separate stimulation events in a single *in vitro* experimental setup. **c**, The mote was operated at 55 mm depth through a fresh porcine specimen. The wireless link traversed approximately 1.5 mm skin, 1 mm fat, 47.5 mm muscle (gastrocnemius and soleus) and intervening connective tissue, and 5 mm gel. **d**, Mote waveform during operation at 55 mm depth through *ex vivo* porcine tissue showing stimulation at 400 μA for 392 μs drove a mock load (purely resistive) of 3.3 kΩ. (a) N = 1 set of two consecutive pulses out of 11 similar demonstrations of slightly differing stimulation parameters. (b) N = 878 total analysed pulse sequences split between stim-applied and no-stim-applied cases. (c) One image of 49 taken, see Fig S12. (d) Representative example pulse out of N = 4 stimulation pulses in this experimental condition.

The StimDust system, while operating in a gel tissue phantom and using an acoustic field with $I_{STPA}$ derated of 551 mW/cm$^2$, achieved: stimulation power of 89 μW (23 nC and 37 nJ per pulse at 2380 Hz), power transfer efficiency from acoustic power at the face of the mote to usable electrical power in the mote of 3.4%, and power transfer efficiency from acoustic power at the external transducer to electrical power in the mote of 0.7% (Supplementary Section 1 and Supplementary Table 1 includes detailed analysis of wireless link efficiencies for operation in a tissue phantom and operation *in vivo*).

The error between specified current output and actual current output across all 8 amplitude levels (50 μA to 400 μA) was 2.0% on average and 8.1% maximum (maximum absolute current error of 5.7 μA). The system achieved up to 2.38 kHz pulse repetition frequency (PRF) while stimulating with a pulse width of 80 μs at 400 μA and minimal compliance limiting. A PRF of 5.05 kHz was achieved with 50 µA current and 80 µs pulse width. Higher frequencies resulted in a $V_{DD}$ droop and compliance-limited current output, though absolute maximum PRF was 16 kHz. Piezo impedance modulation via the $M_{mod}$ switch for uplink data modulation showed a drastic amplitude change in $V_{Pz+}$ (Fig. 4a,d, Fig 6b,c). At the external transceiver, the actual uplink bit error rate (Fig. 4b) was 0 out of 878 and the estimated Gaussian fit bit error rate was $9.9 \times 10^{-8}$ (with the threshold set to minimize bit error rate).

The system operated wirelessly at 55 mm depth through fresh *ex vivo* porcine hindlimb tissue, demonstrating performance at moderate-depth in real tissue with several heterogeneous layers (Fig. 4c,d). The external transceiver generated an acoustic field of 451 mW/cm$^2$ derated $I_{SPTA}$, supplying a total of 19.4 mW $P_{acoustic}$ at the skin which passed through approximately 1.5 mm skin, 1 mm fat and connective tissue, 47.5 mm muscle and connective tissue, and 5 mm gel. The mote generated 65 µW $P_{electrical}$ at VDD, yielding a power transfer efficiency from acoustic power at the skin to electrical power in the mote of 0.33%. Stimulation was demonstrated at all parameters up to 400 µA current and 1500 µs duration.

## Effect of acoustic link alignment

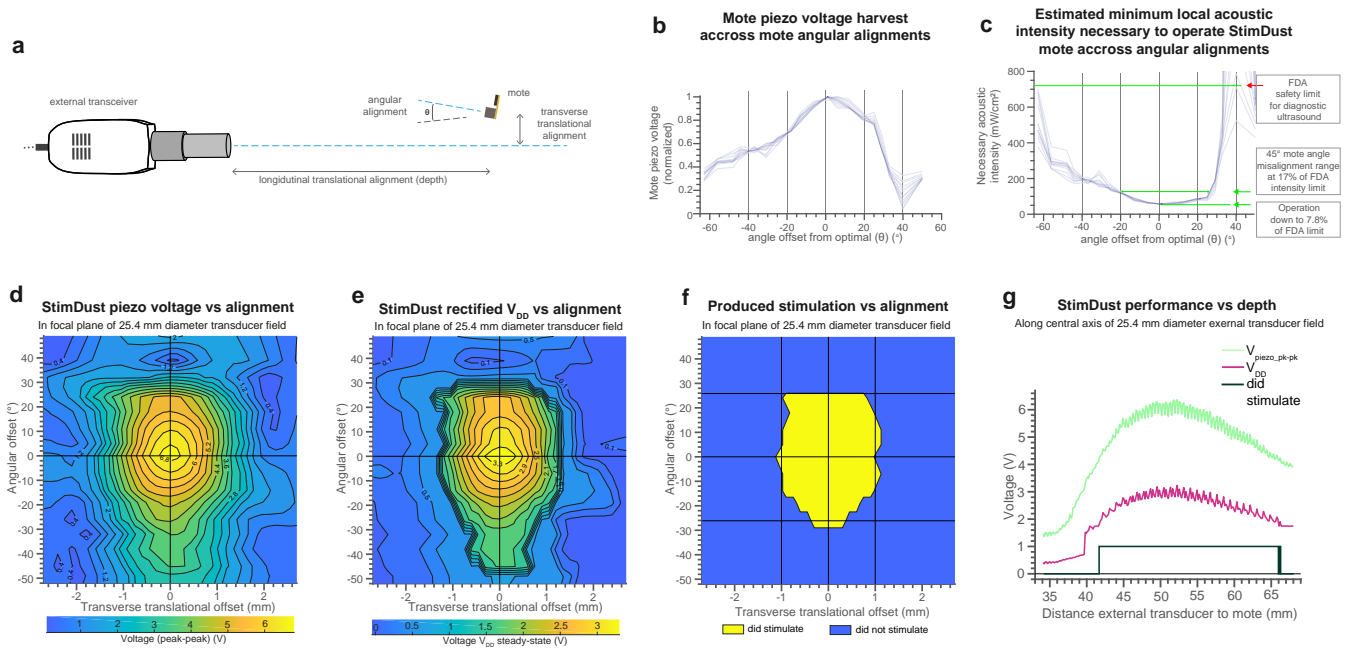

**Fig. 5 | The StimDust mote operated at an intensity 7.8% of the ($I_{SPTA}$ derated) safety limit for diagnostic ultrasound, and with a wide mote angle range.** With an acoustic link traversing an ultrasound gel tissue phantom and using a 25.4 mm diameter external transducer, the StimDust system operated within a 50 degree angular and 2.2 mm transverse translational window at the focal plane, and up to 66 mm depth. **a,** Diagram of misalignment parameters. **b,c,** Mote performance versus angular alignment in arbitrary acoustic field. **b,** Mote piezo peak-peak voltage versus angle of the mote within the acoustic field, normalized to the optimal angle. Multiple lines represent 13 transverse translational offsets of the mote within the full-width-half-maximum region of the incident acoustic field, showing that angular misalignment effects were mostly independent of translational misalignment. The asymmetry was due to the non-centred position of the piezo on the mote. **c,** Given the minimum acoustic intensity needed to operate motes at an optimal mote angle (56 mW/cm$^2$), that for non-optimal angles was estimated. **d**, **e**, **f**, A mote was operated within a fixed external transducer acoustic field (25.4 mm diameter transducer, 25.0 V$_{pk-pk}$ transmit) while its position and orientation were scanned across transverse offset and angular offset relative to the mote central axis. Depth was held constant at 48 mm. **d**, Mote piezo harvest voltage. **e**, Mote regulated voltage. **f**, Region where stimulation was produced. **g**, A mote was operated while its depth was scanned relative to a fixed external transducer acoustic field (25.4 mm diameter transducer, 31.5 V$_{pk-pk}$ transmit), yielding a viable region in depth.

The use of an ultrasonic link instead of an electromagnetic one was crucial for achieving relatively high link efficiencies at sub-mm transducer sizes and cm-depths in tissue. An important consideration for the practical deployment of such a system is understanding the effect of acoustic link alignment on the operation of the mote.

In order to tolerate link misalignment, the implant required a certain minimum power to function; this depended on the local acoustic intensity of the ultrasound field at the mote position and the angular alignment between the thickness-axis of the mote piezo and the local acoustic wave vector direction. To this end, the external transceiver produced an ultrasound field such that the mote received sufficient acoustic intensity within that field. That field was limited, though, as the global maximum acoustic intensity ($I_{SPTA}$ derated) of the field must be below the 720 mW/cm$^2$ the safety limit for non-therapeutic diagnostic ultrasound[40]. $I_{SPTA}$ derated is the time-average acoustic intensity (acoustic power flux) at the spatial point in the field of maximum value when the field is derated by the FDA standard soft-tissue acoustic attenuation of 0.3 dB/cm/MHz.

The minimum acoustic intensity necessary to fully operate StimDust was 56 mW/cm$^2$, 7.8% of the safety limit ($I_{SPTA}$ derated) for diagnostic ultrasound of 720 mW/cm$^2$ [40]. The mechanical index was 0.031 (pressure amplitude of 41.5 kPa), 1.9 % of the safety limit. This more than twelve-fold margin allowed operation of the devices while tolerating link misalignment (Fig 5) and while keeping the ultrasound intensity at safe levels at all points in the acoustic field. The mote had a wide acceptance angle, with piezo voltage harvest varying only between 69% and 100% of maximum over a 45° angular misalignment range of the mote relative to the acoustic field (Fig. 5b). This corresponded to a 45° mote angular misalignment range within which the local acoustic intensity necessary to operate the mote was at most 122 mW/cm$^2$, 17% of the safety limit, and a 75° mote angular misalignment range at with minimum acoustic intensity of was at most 259 mW/cm$^2$, 36% of the limit (Fig 5c). This indicated that, for example, a mote would safely operate anywhere in the body where the local ultrasound intensity is at least 122 mW/cm$^2$, the mote is within a 45° angular alignment range, and the global maximum field intensity is less than 720 mW/cm$^2$, and. The wide mote angular alignment range was critical because, while it was relatively easy to adjust the ultrasound field generated by the external transceiver after mote implantation by moving the external transceiver (future versions could use beamforming) or swapping out external transducers, adjusting the orientation of the mote itself after implantation was difficult.

This wide window of operation of the mote also enabled flexibility when generating the external field. In this work, two off-the-shelf single-element focused transducers were alternately used to generate acoustic field shapes optimized for different depths (acoustic field characterization in Fig. S3 a,b). Using a 25.4 mm external transducer driven so that $I_{SPTA}$ derated was 584 W/cm$^2$, the StimDust system could reliably operate when the mote was within a depth range of 42 to 66 mm (Fig. 5g) (up to 70 mm maximum operational depth observed in other experiments). At $I_{STPA}$ derated of 466 mW/cm$^2$, the system functioned when the mote angular orientation was within a 55° range around the central axis, and operated when the external beam was centred within a 2.2 mm transverse translational range (as measured at the focal plane) around the mote position (Fig. 5d-f).

**Stimulation *in vivo***

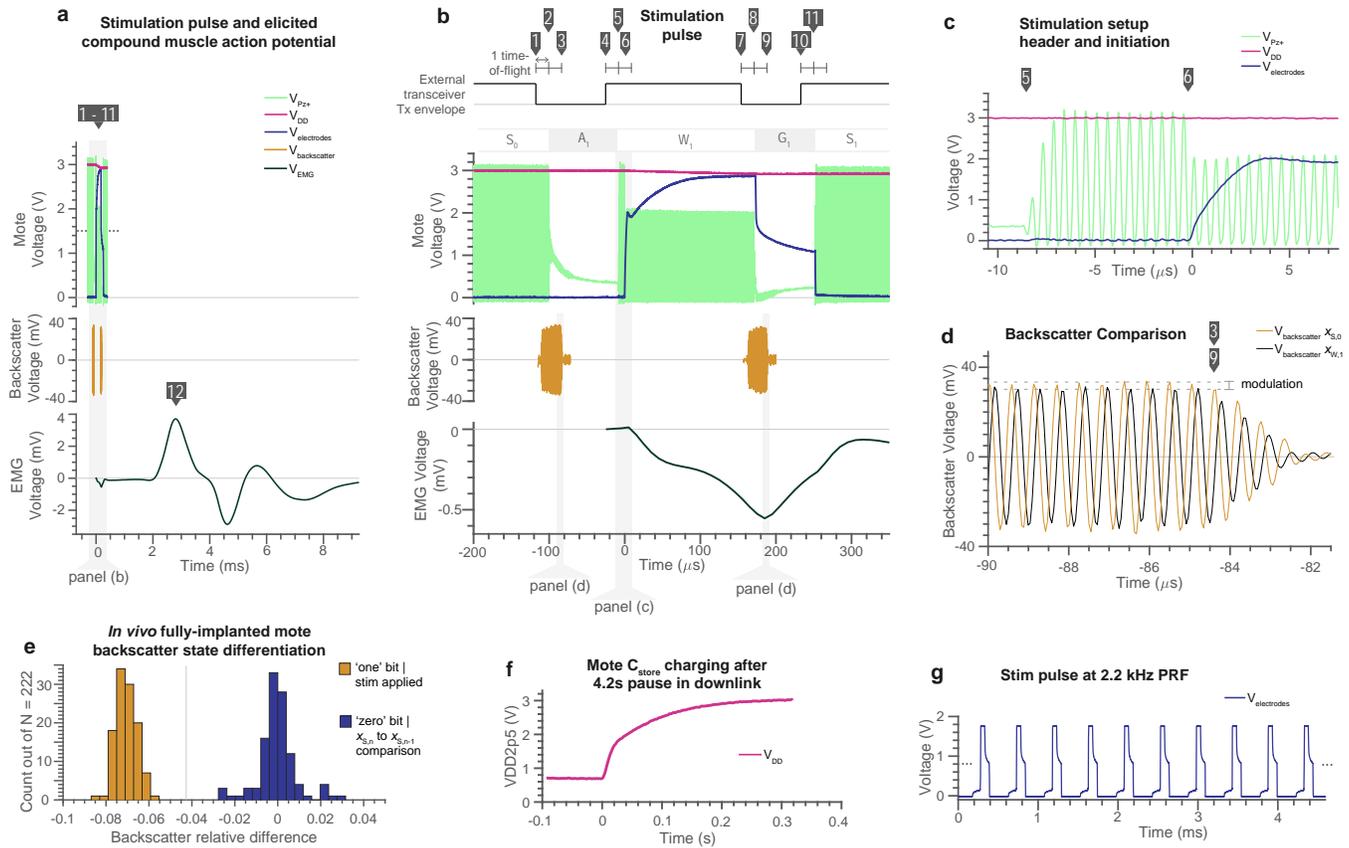

**Fig. 6 | *In vivo* performance: mote, backscatter, and evoked neural response waveforms for fully-implanted mote. a**, A stimulation pulse from the mote (a top) elicited a compound muscle action potential (a bottom). **b**, Detail of stimulation pulse signals. The downlink code regions $A_1$, $W_1$, and $G_1$ specify the stimulation pulse. During $S_0$ the ultrasound envelope was high ($V_{pz+}$), the charge in the energy storage capacitor was topped off ($V_{DD}$), and the electrodes were shorted ($V_{elec}$). **[1]** The external interrogator set the downlink ultrasound envelope from on to off, and after a short delay, began to capture the backscatter signal $x_{S,0}$ from the mote from the $S_0$ period ((b), middle). **[2]** One acoustic time-of-flight (approximately 16.9 µs in this trial) later the mote saw a transition from $S_0$ to $A_1$, beginning the first ultrasound-free interval and initiating the current amplitude TDC decoder. **[3]** One time-of-flight after the mote $S_0$ to $A_1$ transition, the primary backscatter from the mote $S_0$ region as seen at the external transducer ceased ((d), orange line). **[4]** The external interrogator set the downlink ultrasound envelope from off to on. One time-of-flight later, at the mote $A_1$ to $W_1$ transition **[5]**, the stimulation amplitude was set, and a header / setup sequence lasted for 16 ultrasound cycles **(c)**, after which **[6]** the stimulation output pulse began and $M_{mod}$ turned on, reducing the amplitude of $V_{Pz+/-}$. During stimulation, the voltage measured at the stimulation electrodes showed an immediate $I_{stim} \cdot R_{soln}$ increase, and then gradually increased due to charging of the double layer capacitance. $V_{DD}$ dropped linearly during this period due to constant current flow through the stimulation electrodes. In addition, there was an electrical

stimulation artifact on the EMG signal during the stimulation event ((b), bottom). **[7]** The external transceiver began the second ultrasound-free interval and after a short delay, began to capture the backscatter signal $x_{W,1}$ from the mote from the $W_1$ period ((b), middle). **[8]** At the mote $W_1$ to $G_1$ transition, the stimulation electrodes were set to high impedance for the interphase gap. The electrode voltage showed an immediate $I_{stim} \cdot R_{soln}$ drop and then a $R_{soln} \cdot C_{dl}$ discharge as the double-layer capacitance discharged. **[9]** One time-of-flight later the primary backscatter seen at the external transceiver from the $W_1$ period ceased ((d), black line). **[11]** At the mote $G_1$ to $S_1$ transition, the stimulation electrodes were shorted to discharge any remaining charge. **[12]** The stimulation pulse produced by this sequence of events evoked the compound muscle action potential seen in (a, bottom). **d,** Detail of backscatter signal captured at external transceiver, showing amplitude modulation. **e,** The two regions from (d) were compared to yield 111 'one' bits, with good separation from 111 'zero' bits (grey line depicts threshold), (see Methods for how 'zero' bits were calculated in this instance). These replications are separate stimulation events in a single animal. **f,** Initial charge-up of the 4 μF $C_{store}$ after the mote had been without power for 4.2 seconds and $V_{DD}$ had drained to 0.7 V. **g,** High pulse repetition frequency in-vivo stimulation. The small electrode voltage rise which precedes the main ohmic rise in each of these pulses is due to the remaining charge on the simulation electrodes from the previous stimulation pulse which was not fully cleared during relatively brief ~230 μs shorting phase. Note: data in panels (a-d) are from a single representative stimulation pulse out of N = 111 stimulation pulses analysed for the experimental condition (115 pulses acquired; see *Methods*), which was conducted during an *in vivo* experiment with a fully implanted mote (animal D). Data in panel (e) are from all 111 stimulation pulses analysed for the experimental condition (all stimulation pulses had the same set of stimulation parameters). Data in panel (f) is a representative example out of N = 2 stimulation pulses which are from the same *in vivo* experiment (Animal D) as the data in panels (a-e) and had the same stimulation parameters and qualitative biological effect, but for which the oscilloscope voltage measurement setup was configured to capture the longer timeframe of $V_{DD}$ charge-up and not the short details of the stimulation pulse. Data in panel (g) is from *in vivo* while taking data with an open surgical site (animal A). The data originates from a dataset of 12 different stimulation parameter sets sweeping pulse repetition frequency from 1 Hz to 3200 Hz where N = 1 for each condition; the data in panel (g) depicts the stimulation waveform at the highest achieved stable stimulation pulse repetition frequency in this experimental condition sweep.

*In vivo* stimulation elicited robust and repeatable compound action potentials. Stimulation in an acute preparation was demonstrated in six animals and data is presented for animals D (Figs. 6, 7, S5), C (Fig. S6), and F (Fig. S6). The motes implanted in animals C and D had overlaying muscle and skin layers replaced and sutured after implantation and were operated with the ultrasonic link traversing these tissue layers (Fig. 6, Fig. S6, Supplementary Video 1). The acoustic field (Supplementary Fig. 3) had a derated $I_{SPPA}$ and $I_{SPTA}$ of 692 mW/cm$^2$ and mechanical index of 0.11, both below the FDA limits of 720 mW/cm$^2$ and 1.9, respectively, for ultrasound diagnostic imaging[40], and lower than thresholds for directly ultrasound-mediated stimulation (Supplementary Section 2). Fig. 6a-d shows all signals of interest during a stimulation pulse while a fully implanted mote was operated through skin and muscle layers. The downlink commands received by the mote, $S_0$, $A_1$, $W_1$, $G_1$, $S_1$, can be seen in the $V_{Pz+}$ trace in figure 6b, top. The conduction delay between stimulation and evoked CMAP of approximately 1.5 ms corresponded to the expected conduction velocity of approximately 25 m/s[52] and distance of approximately 30 mm. Through-tissue backscatter uplink performance (Fig. 6d, e) had an actual bit error rate of 0 out of 222 (from 111 pulses; see Methods) and a Gaussian fit estimated bit error rate of $1.5 \times 10^{-7}$. This indicated a robust 1-bit data uplink that could reliably report the success of each pulse to the user.

## Stimulation parameters control physiological response

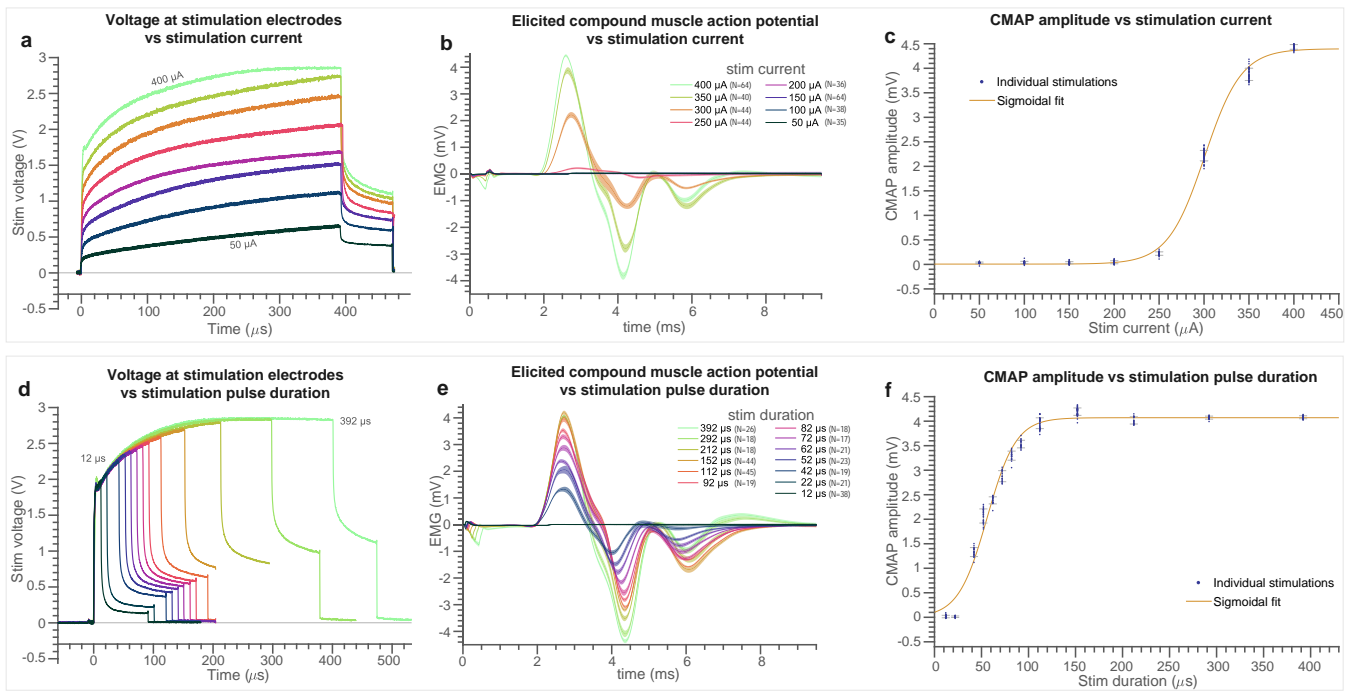

**Fig. 7 | Precise control of evoked neural response achieved through varying stimulation current or stimulation pulse width. [Current-control]**: Stimulation current was varied from 50 μA to 400 μA with stimulation pulse width held at 392 μs. **a,** Stimulation electrode voltage shows an approximately linear trend with stimulation current (N = 1 stimulation pulse waveform collected for each condition). **b**, CMAP (compound muscle action potential) waveforms show activation for 250 μA to 400 μA. For each condition, the coloured line represents the trial-average CMAP voltage at each point in time, the width of the shaded error region is ± 1 s.d. (standard deviation), and N is the number of pulse events in each condition. The pulse repetition rate of approximately 1/3 Hz yielded essentially independent biological responses to each pulse, though there could have been small non-independent effects due to muscle fatigue. **c**, CMAP amplitude vs. stim current shows a typical sigmoidal recruitment curve. Each point is a single CMAP baseline-to-peak amplitude, the line is a sigmoidal fit and error bars are ± 1 s.d. of the CMAPs in each stim current condition. The sample size for each condition in panel c is the same as that for each condition in panel b. **[Pulse-width-control]**: Stimulation pulse width was varied from 12 μs to 392 μs with stimulation current held at 400 μA. **d,** Stimulation electrode voltage shows an approximately linear trend with stimulation current (N = 1 stimulation pulse waveform collected for each condition). **e**, CMAP waveforms show activation from 42 μs to 392 μs. For each condition, the line represents the trial-average CMAP voltage at each point in time, and the width of the shaded error region is ± 1 s.d., and N is the number of pulse events in each condition. This panel shows that the duration of the electrical artefact increased with increasing stimulation pulse width, but the time-course of the CMAP did not appreciably change, with only amplitude differing. **f**, CMAP amplitude vs. stim pulse width shows a typical sigmoidal recruitment curve with little or no evoked response below 22 μs pulse width and saturation at approximately 140 μs pulse width. Each point is a single CMAP baseline-to-peak amplitude, the line is a sigmoidal fit and error bars are ± 1 s.d. of the CMAPs in each stim duration condition. The sample size for each condition in panel e is the same as that for each condition in panel f. Note: This data was taken with an open surgical site. All replications in this figure are separate stimulation events in a single animal.

*In vivo* EMG response was controlled by varying stimulation current from 50 µA to 400 µA with fixed 392 µs pulse width (Fig. 7a-c). The low standard deviation of the CMAP waveforms (line width in Fig. 7a, d) indicated highly repeatable stimulation and biological response across pulses. The recruitment curve as a function of stimulation current (Fig. 7c) showed a threshold of activation between 200 µA and 250 µA and saturation between 350 µA and 400 µA. The CMAP amplitude saturated at 4.4 mV. In vivo EMG response was also controlled by varying stimulation pulse width with fixed 400 µA current (Fig. 7d-f). The CMAP amplitude response showed the duration-threshold of elicitation of action potential at 400 µA current to be approximately 26 µs, with response saturation around 150 µs and maximum CMAP amplitude of 4.2 mV.

A second set of current-control and pulse-width-control sweeps were performed two hours after the first (Supplementary Fig. 5). Wireless link performance, communications, and stimulation output were unchanged. The biological response was very similar but showed recruitment curves shifted slightly left, indicating a 15% reduced threshold of recruitment, and 11% lower CMAP amplitude saturation. These slight differences were potentially due to impedance changes at the electrodes and excitability in the nerve as the tissue began to responds to the implant, muscle fatigue (though pulse repetition frequency was low), or changes in the plane of anaesthesia.

Measurement of CMAPs from multiple innervated muscles in animal F (Fig S6) (the medial and lateral heads of the gastrocnemius, and the tibialis cranialis) show strong co-activation of muscle groups, indicating that stimulation was delivered to multiple fascicles in the nerve.

*In vivo* stimulation across pulse repetition frequencies from 0.22 Hz to 2.22 kHz yielded four regimes of muscle activation. Pulse width and current were fixed at 52 µs and 400 µA. By visually observing muscle twitch response, we qualitatively reported single twitches at 0.25, 0.33, 1, 2, 5, and 10 Hz, unfused tetanus at 32 Hz, and fused tetanus at 100 and 320 Hz. This matched expected muscle responses for rat gastrocnemius[53]. At 2 kHz and 2.22 kHz (Fig. 6g), we observed a single initial twitch and then apparent relaxation. While we did not measure muscle force in this study and thus could not positively confirm nerve block, nerve block has been reported at 2 kHz stimulation frequency and 1 V drive voltage (approximately 666 µA)[54]. Therefore, while not conclusive, these observations were consistent with the system performing nerve block.

With stimulation and return electrodes each sized 550 µm x 550 µm, the voltage drop across the spreading resistance, also known as access voltage, ($I_{stim} \cdot R_{soln} \approx 1.7$ V) accounted for a majority of the stimulator headroom. This allocated approximately 0.6 V of headroom across each electrode interface and kept the electrodes safely operating within the electrochemical water window of PEDOT:PSS (-0.6 V to 0.8 V). While asymmetric electrodes could utilize the entirety of the water window during the stimulation phase by preferentially charging stimulation electrode to a higher voltage (return electrode larger than stimulation electrode), we opted to use symmetric electrodes since the voltage on the stimulation electrode is inverted during the shorting phase. The maximum combined interfacial potential was considered to be the voltage across the stimulation electrodes 20 µs into the interphase gap[55]. This value was below 1.23 V under all stimulation pulse conditions less than 400 µA current and 392 µs pulse duration (the maximum conditions tested). At this maximum corner-case, the maximum combined interfacial potential was between 1.31 and 1.36 V. This is slightly above the maximum range, but this corner-case condition can be avoided while still accessing the full recruitment curves. And minor overpotential at or above this corner-case would be very small and only last a few tens of µs. Furthermore, the two electrodes were in close proximity, enabling rapid aqueous diffusion and recombination of any potential reaction products above this stimulation level. Finally, an updated design of StimDust could increase electrode area marginally to mitigate any overpotential.

## Discussion

Our results indicated that repeatable, controlled compound action potentials can be elicited by the StimDust system in an acute preparation. Grading the physiological response through controlling stimulation current, pulse-width, and pulse repetition frequency demonstrated efficacy for a wide range of stimulation techniques and is an important component of safety. Furthermore, miniaturization of the precision stimulator enabled placement directly at the stimulation site *on* the sciatic nerve.

**Table 1:** State-of-the-art comparisons table

|  | TBioCAS '18 [38] | TCAS II '13 [37] | PLOS ONE '17 [28] | Sci. Reports '18 [31] | Front. Neuro. '17 [29] | TBioCAS '18 [56] | **StimDust** This Work |
|---|---|---|---|---|---|---|---|
| Stim. type | current | current | voltage | current | voltage | current | **current** |
| Link | ultrasound | ultrasound | RF | 4 coil inductive | RF | RF | **ultrasound** |
| $f_{carrier}$ [MHz] | 1.3 | 1 | 2400 | 13.56 | 10 | 1300 | **1.85** |
| Uplink | none | none | none | 433 MHz OOK | none | none | **backscatter** |
| Process [nm] | 180HV | 350 | discrete | 130 | discrete | 130 | **65** |
| Waveform configuration | digital bitstream | digital bitstream | received power | digital bitstream | received power | received power | **envelope detection** |
| $F_{Stim}$ [Hz] / resol. [bit] | 0 - 60 | 60 - 265 /2b | 0 - 25* | 13-414 /5b | 2* | 0 - 3* | **0 - 5k**** |
| Stim start | enable signal | -- | received power | stim command | received power | received power | **enable signal** |
| $T_{Pulse}$ resolution [us] | 14 / 5b | 200 / fixed | continuous | 9.5 / 5b | continuous | continuous | **continuous** |
| $I_{Stim,max}$ [µA] / resol. [bit] | 5000 / 8b | 640 / 5b | undefined | 1860 / 5b | undefined | 25 / fixed | **400 / 3b** |
| $T_{gap}$ resolution [µs] | 1/2 $T_{pulse}$ | -- | N/A | 9.5 / 5b | N/A | N/A | **continuous** |
| Charge balance | biphasic/gnd | dc-blocking cap | none | 3b biphasic correction | none | none | **passive recharge** |
| Compliance voltage (V) | 15 | 3.3 | N/A | 2 | N/A | 1.2 | **3** |
| Animal model | ex-vivo frog sciatic | rat abdomen | pig vagus | rat sciatic | rat sciatic | ex-vivo rat sciatic | **rat sciatic** |
| Fully-implanted | no | no | yes | yes | (?; surgical site closed?) | no | **yes** |
| in-vitro depth$_{tx-rx}$ [mm] | 105 (castor oil) | -- | -- | -- | -- | 5 | **70 (gel), 55 (muscle)** |
| in-vivo depth$_{tx-rx}$ [mm] | -- | 50 | 15 | -- | 75 (through air) | -- | **18** |
| Mass [mg] | 78 | -- | 300 | 2800 | -- | -- | **6.4** |
| Circuit η ($P_{stim}/P_{in}$) [%] | <50 | -- | 20 | -- | -- | -- | **82** |
| Volume [mm³] | 39 | 1020 | ~17 to 41 | 2250 | 0.45‡ | -- | **1.7 ‡** |
| Electrodes [mm] | 0.5Ø x 2.0 Pt cylinder | -- | 2.7Ø x 1.0 Pt cuff | cuff | 0.3Ø Pt disk | 0.125Ø SS disk on cuff | **0.55 PEDOT square** |
| Implant encapsulation | PDMS | -- | epoxy | epoxy & PDMS | epoxy | -- | **parylene** |
| FOM (Depth/Volume) | 2.69 | 0.05 | ~0.36 to 0.88 | -- | -- | -- | **32 §** |
| FOM (Volume/Efficiency) | 0.78 | -- | 2.03 | -- | -- | -- | **0.02** |

*demonstrated, range not provided  **up to 16kHz in burst
†estimated  ‡includes stim electrodes but not cuff

Table 1 summarizes the performance of StimDust and compares it with state-of-the-art implantable wireless neural stimulators[38,37,28,31,29,56]. Wireless ultrasonic power and data delivery enabled implant operation at a depth of 70 mm in ultrasound gel. Additionally, direct control of the stimulation waveform by the received ultrasound boasted several advantages over state-of-the-art. Compared to systems that program motes via a bitstream, StimDust achieved sub-µs (1/$f_{carrier}$) temporal resolution with a larger dynamic range for pulse width, interphase gap, and PRF because the complexity was pushed to the external transceiver, which had fewer computational and power constraints. Stimulators which used only passive components[28,29] used a direct control scheme as well; however, the stimulation intensity was dictated by the received power and the output was voltage stimulation which is sensitive to electrode properties. Furthermore, they did not feature an uplink whereas StimDust included a wireless uplink via ultrasonic backscatter for safety monitoring and alignment. Active rectification increased efficiency and established a ground reference potential, enabling StimDust to use passive recharge to save power relative to biphasic stimulation while clearing residual stimulation charge on the electrodes, which other direct control systems[28,29] were unable to do. Using an enable signal, or stimulation start command, allowed the mote to stimulate with low latency, which is crucial for closed-loop experimentation.

In addition to highly programmable stimulation, StimDust was also the smallest neural stimulator with precise current-control output. The efficient integration and co-design of the piezoceramic, Stim IC, storage capacitor, and interface electrodes enabled miniaturization of the mote down to 1.7 mm$^3$ in volume and 6.4 mg. This represents >10x improvement in depth/volume figure of merit (FoM) and >25x improvement in volume/efficiency FoM relative to other state-of-the-art precision neural stimulators. A further-miniaturized revised StimDust design yielded a mote volume of 0.65 mm$^3$ and expected depth/volume FoM of 78 mm$^{-2}$, a >29x improvement over state-of-the-art.

The use of ultrasound as the wireless power carrier enabled direct voltage harvest from the mote piezo of up to ~7 V$_{pk-pk}$ and ~3.5 V rectified under safe ultrasound intensity levels, which is sufficient for stimulation with low-impedance electrodes. We chose to take advantage of this to optimize for high power-efficiency rectification. While operating at clinically relevant depths of multiple centimetres, this enabled our high PRF (and high total P$_{stimulation}$), which is important for some clinical use cases such as nerve block. The acoustic-to-electrical power conversion could have alternatively been optimized for voltage efficiency rather than power efficiency which would have enabled greater depth of operation or a thinner piezo (both of which reduce mote piezo voltage) at the expense of maximum PRF and P$_{stimulation}$.

This approach compares favourably to existing ultrasound-based approaches in its small size, precise stimulation, and controllability; it is smaller than any other wireless neural stimulator capable of current-mode stimulation. It compares favourably to electromagnetic-based wireless implantables in the power that can be harvested in a very small implant at moderate depths in the body. If, for a particular application, these constraints are of low priority, then a radiofrequency EM approach may have a simpler user experience, as such approaches can typically tolerate an air gap between the external transceiver and the body[28], whereas an ultrasonic link requires acoustic contact at the transceiver-body interface (though a link requiring contact may improve security). In relation to traditional battery-powered implantable stimulators, which are large and rely on occasional reprograming, the ultrasound approach compares favourably due to its small size, completely untethered design, and low-latency downlink and uplink. If size and controllability are not priorities, then a traditional stimulator has the advantage of not requiring an externally-worn transceiver.

**Robustness to acoustic link alignment**

Link alignment measurements showed that for general use of the system, operation was not very sensitive to depth alignment, and could be fine-tuned by changing the stand-off distance between the mote and the external transceiver with added ultrasound gel or a gel pad, or by swapping external transducers with differing focal lengths. Operation was also not very sensitive to mote angular alignment, which was controlled by surgical placement of the mote with approximately aligned orientation for the anticipated location of the external transceiver. Operation was somewhat sensitive to lateral misalignment of the mote relative to the central axis of the acoustic field, which was related to both the position and angular orientation of the external transceiver relative to the mote, and was dependent on the particular off-the-shelf external transducers utilized for this work. The demonstrated transverse alignment range of 2.2 mm at the focal plane was limited by the field shape generated by the external transducer, and is expected to be sufficient for applications that do not experience significant motion, such as near moving joints.

This acoustic field shape was determined by an acoustic lens built into the external transducer, which was limited in options by commercial availability. Exploring the design space of transducer diameter and lens shape can change the locations in the body where the acoustic field, combined with the mote power and angular misalignment properties, enable the device to function. Several acoustic aperture and lens design techniques exist [57, 58, 59] which can sculpt the generated acoustic field according to specified constraints.

Using simple spherical or parabolic lens design, the beam width scales as focal length / aperture. Wider beam widths could be employed to reduce the sensitivity to transverse misalignment. A longer focal length and wider aperture could be used to improve maximum depth of operation. Such longer focal-length transducers[60,61] would be able to operate a mote at over 100 mm of depth, though operation at such depths would have to be balanced against the increase in transverse link misalignment caused by a given angular misalignment of the external transceiver. Generalizing beyond spherical and parabolic lenses allows for an acoustic field region of relatively uniform intensity which extends relatively far from the central axis (a 'wide beam'), enabling acoustic link operation under greater transverse translational misalignment. An external transducer with such a designed lens, for example, could generate an acoustic field (with $I_{STPA}$ limited to the safety limit with standard derating) which has a region 11 mm wide, from 53 to 57 mm depth, in which the acoustic intensity is everywhere greater than 122 mW/cm$^2$, sufficient to operate a mote within a 45 degree range of mote angular alignment. In conjunction with a wider beam, active measures could be employed, such as beamforming from an array of elements[62], or beam translation from use of different subsets of a larger and wider element array.

With these changes, it is possible to make an external transducer field with a power delivery region which will open up the transverse translational misalignment window and the external transceiver angular misalignment window, enabling a greater range of real-world use with reduced requirements for link alignment stability.

**The StimDust technology can be extended to suit various applications**

The demonstration of the StimDust stimulator in this work was carried out on a peripheral nerve while utilizing a simple cuff-type nerve attachment most similar to a split-cylinder style cuff. Translation to other stimulation targets and devices with modified morphologies is a ready possibility. For example, with modification to electrode feed-throughs or vias, the StimDust simulator could be attached to a self-sizing spiral cuff[63,19] or helix-style cuff[64,65] which have shown safer levels nerve pressure over a wider range of nerve diameters, increasing safety against nerve damage. Furthermore, if fibre-tract specificity is critical for the clinical use case, the StimDust stimulator could be paired with cuff and electrode styles which achieve specificity, such as a multi-contact cylindrical cuff device[19], a FINE-style cuff (flat interface nerve electrode)[66,67], sieve electrodes[68] or intrafascicular electrodes[69,70]. While no nerve pressure or strain measurements were taken as part of this study, it is expected that an implementation of StimDust with any cuff style that is completely wireless, tethereless, and leadless, will reduce the incidence of lead-induced strain on the nerve, especially in areas of high motion such as joints.

For certain applications, such as producing differentiated percepts in artificial sensory stimulation[8] or increasing input dimensionality[15], stimulating at multiple sites is an important capability. This could be achieved with StimDust with multiple independent motes or multiple stimulators attached to a single cuff-like device. These motes would be operated with independent ultrasonic links, generated by independent external transceivers or a single external transceiver with an array transducer. In the present design, the ultrasonic link is an approximately gaussian beam with a full-width-half-maximum of 1.6 mm in the focal plane (Fig S3). Thus, two stimulators separated by approximately 2 mm or more would be able to be operated independently with separate acoustic links and external transceivers. Alternatively, a revised StimDust design could incorporate multiplexing into the Stim IC without additional external components. This requires a modification of the communication protocol to select the source and sink electrodes for the stimulation pulse. This multiplexing strategy when combined with the previously multi-mote method, drastically increases the number of stimulation sites.

An important use case for a wireless neural stimulator is placing the mote in the brain for use in DBS and brain-machine interfacing. The diminished size, potential for ultra-minimally invasive delivery, and totally wireless operation opens up possibilities such as wireless multi-site deep brain stimulation and cortical stimulation for proprioceptive feedback. If a small and wireless stimulator is shown to have lower risk in chronic clinical settings, then the risk-threshold for neurostimulation will be lowered, enabling it's use in a wider set of indications such as certain psychiatric conditions[12,13]. Intracranial use-cases pose a new challenge, though: establishing an ultrasonic power and communications link through the skull is challenging due to the greater attenuation in skull than in soft tissues. Transcranial ultrasound has been shown to safely achieve ultrasonic power levels in the brain of 159 mW/cm$^2$ at 2 MHz and a focal depth of 4.5 cm[71,72]. StimDust is able to operate at a power level (56 mW/cm$^2$) and frequency (1.85 MHz) lower than this, indicating potential applicability to transcranial and deep-brain operation.

**Clinical Translation**

In order to clinically translate StimDust for chronic use cases, long-term performance over a meaningful treatment period must be demonstrated. To achieve device lifetimes longer than several months to a few years, the parylene encapsulation used for StimDust in this work can be replaced with device packaging or encapsulation with greater hermeticity. Ceramic and metallic packages have been shown to provide excellent hermiticity and can scale down to millimetres[73,74,75]. Recently, a millimetre-scale alumina-titanium package was used to house an acoustically powered IC, and both ultrasonic powering and backscatter communication were demonstrated through the sealed package[76]. Thus, the StimDust stimulator could be packaged using proven processes and materials While using such readily-implementable packaging techniques would increase the volume of the stimulator relative to a thin-film conformal encapsulation, optimization of the layout of components of the stimulator can help mitigate this volume growth. A more aggressively scaled mote of 0.65 mm$^3$ in volume (before packaging) was designed (Supplementary Fig. 2) which used the same system architecture, piezo size, and fabrication techniques as the 1.7 mm$^3$ mote design experiments reported in this paper.

In addition, inert conformal thin-film coatings are a promising method of providing hermetic encapsulation without significantly increasing the volume of the device. 1-µm-thick silicon dioxide device encapsulation[77] has been shown to provide a barrier for electronics in the biological environment for a projected 70 years. Silicon carbide encapsulation[78] has shown even greater longevity, and has an encapsulation thickness of 550 nm to 1000 nm. Silicon carbide can also be doped to create conductive electrode sites, obviating electrode pass-throughs and material interfaces for crack nucleation or water vapor permeation. It is important to note that non-toxic piezoelectric materials, such as barium titanate[79] can be used in place of PZT, with a reduction in acoustic power harvesting of approximately 27%[80].

The StimDust design in this work used PEDOT plating on the gold electrodes to improve charge injection capacity and enable small electrodes[81]. PEDOT coatings on smooth metallic substrates can delaminate after approximately a year in the biological environment[82]. To achieve comparable stimulation performance with clinically-proven long-lasting electrode materials, StimDust can use electrodes with larger area. The reported device had an electrode area of 0.303 mm$^2$ and demonstrated a charge injection limit of 137 nC ph$^{-1}$ *in vivo*. The electrode area can be increased to 2.5 mm$^2$ within the existing footprint of the StimDust substrate board, and electrode area can be increased by to 5.0 mm$^2$ if the electrodes use the inside surface of a 1.2 mm diameter cuff. A 2.5 mm$^2$ platinum or platinum-iridium electrode has an estimated charge injection limit in the range of 415 nC to 2850 nC [83,84], substantially more than what was needed during the *in vivo* experiments performed with StimDust. Clinically approved stimulators use smooth platinum-iridium electrodes of approximately this area: the Cyberonics Helix electrode, used clinically for vagus nerve stimulation, has electrodes with area estimated to be 3.1 mm$^2$ each[85,86] and the Boston Scientific directional DBS lead utilizes platinum-iridium electrodes of approximately 2 mm$^2$ area[87].

Various techniques and materials have shown that higher charge injection capacity electrode materials can be fabricated that have good longevity. This includes Electrodeposited Platinum-Iridium Coating (EP-IC)[88] which is a mechanically robust coating that has demonstrated at least 180-day lifetime in vivo and 10-years-equivalent in vitro testing. Platinum black and platinum grass[89] show lower impedance than smooth platinum. Use of PEDOT in a chronic device may be possible by roughening of the metallic layer that PEDOT adheres to[90,91].

Scar tissue can create a sheath of fibrous tissue around the implant. Fibrous tissue, such as cartilage and tendon, can have acoustic impedances slightly higher than that of water and most other tissues[92] leading to an acoustic boundary and a reduction in mote power of approximately 0.23%, a small to negligible impact on the performance of the ultrasonic link. The fibrous encapsulation can also have an effect on the impedance of the tissue. Pt-Ir cuff electrodes measured over 70 days on the cat sciatic nerve[93] have been shown to have a 1000-Hz impedance stabilize at approximately 150% that of its initial value. Another study, measured the resistivity of fibrous encapsulation in cats[94] and reported that implanted platinum electrodes showed a tissue resistivity drop during days 0 to 4 post-implantation followed by a rise to 60% of its initial value. This indicates that StimDust should be able to perform safe and efficacious stimulation even after a drift of ± 50% from the nominal impedance at implantation. The potential for a 50% increase in tissue impedance can be accommodated by the presented StimDust design by increasing the electrode area.

A clinically translated StimDust device with these longevity solutions is predicted to have *in vivo* durability that exceeds that of current clinical devices due to the absence of critical failure modes. Traditional implantable neural stimulator failures are dominated by lead failures, reported to occur in 7.6% [95], 2.9% [96], and 1% [97] of implantations. This can occur through fatigue failure of bending components or at the lead feedthrough to the stimulator[98]. Leaded simulators are typically implanted as separate components and rely on a lead-feedthrough connection that is made intraoperatively. StimDust can be hermetically sealed and quality-controlled in manufacturing, and is implanted as a single monolithic device.

## Outlook

In conclusion, we have introduced the smallest wireless neural stimulator that can deliver current-controlled stimulation. The current-controlled stimulation demonstrates that there does not need to be a trade-off between safe, predictable stimulation and highly miniaturized stimulator size. Compact mote integration and programmable stimulation was enabled by a custom wireless ultrasound protocol and an energy efficient integrated circuit. We demonstrated its use acutely implanted in anesthetized rats to controllably elicit compound action potentials across a range of clinically relevant stimulation regimes. Taken together, this work shows a practical demonstration of a stimulator for the PNS with potential applicability to the CNS. We anticipate the introduction of a class of neural stimulators that are small enough for many to be implanted with relatively low risk to the patient, and which open up powerful therapeutic and neural interface techniques.

## Methods

**Design, fabrication, and assembly of the StimDust motes:** The StimDust motes were built on a 100 µm thick, polyimide, flexible PCB (Altaflex, Santa Clara, CA). Polyimide substrates were preferable to FR-4 substrates due to their higher thermal budget, superior chemical compatibility with organic solvents, and higher flexibility of the substrate, facilitating device testing and *in vivo* implantation.

Stimulation electrodes were located on the underside of the PCB. The electrodes had 0.55 mm x 0.55 mm dimensions with 1 mm centre-to-centre spacing. To prevent compliance limiting when using the full current range of the chip, the electrode impedances were designed to be < 6.25 kΩ at the stimulation pulse frequency (~2.5 kHz for a pulse width of 200 µs – the lowest frequency used and therefore highest impedance). The impedance of the gold electrodes on the PCB was lowered by electroplating poly(3,4-ethylenedioxythiophene) polystyrene sulfonate (PEDOT:PSS) using a NanoZ impedance testing and electroplating device (White Matter LLC, Mercer Island, WA). The nominal plating current was approximately 4 µA/mm² for 60 s in a plating solution of 1:1 ratio 0.02 M EDOT and 0.2 M PSS. In recent work, PEDOT:PSS has demonstrated a charge injection limit 15x that of PtIr and IrOx electrodes and >3x reduction in the voltage transient[81]. The impedance of the gold stimulation electrodes was characterized in 1x phosphate buffered saline (PBS) using a NanoZ before and after PEDOT plating and again after full assembly of the motes. One electrode served as the active electrode while the other served as the reference electrode. A frequency sweep between 1 Hz and 5.02 kHz was performed and the impedance data collected (Fig S8g), yielding an impedance of approximately 2 kΩ at 2 kHz[42], down from approximately 9 kΩ before PEDOT plating. The electrode stimulation voltages were characterized in PBS with relevant pulse currents and durations both before and after PEDOT plating (Fig S8a-f). The charge injection capacity (measured across both electrodes) of the gold electrodes was $19 \pm 5.1$ µC · cm$^{-2}$ · ph$^{-1}$ (N=3) and of the PEDOT electrodes was $188 \pm 61$ µC · cm$^{-2}$ · ph$^{-1}$ (N=3). The charge injection capacity of the PEDOT electrodes adjacent to the sciatic nerve *in vivo* was 45 µC · cm$^{-2}$ · ph$^{-1}$ (N=1).

Following electroplating, the IC was attached to the PCB using a conductive silver epoxy (H20E, Epoxy Technology Inc., Billerica, MA) (Fig. 3b) and cured at 150 °C for 10 minutes. The IC and PCB were then cleaned in acetone, followed by isopropyl alcohol and DI water to clean the device prior to wirebonding. The IC was wirebonded to the PCB using an ultrasonic wedge wirebonder (7400B, West Bond, Scotts Valley, CA) and potted in UV-curable epoxy (AA 3526, Loctite, Düsseldorf, Germany) to protect the wirebonds.

Next, a 750 µm x 750 µm x 750 µm lead-zirconate titanate (PZT) cube (841, APC Int., Mackeyville, PA) and an 0201 surface mount capacitor for charge storage were attached to the PCB using silver epoxy and cured at 150 °C for 10 minutes. PZT was chosen for its high electromechanical coupling constant and relatively low loss tangent enabling good voltage and power harvest in semi-continuous-wave applications. Barium titanate piezoceramic could also be used for improved biocompatibility, but with a slight reduction in electromechanical coupling coefficient. The PZT coupon was diced from a 750 µm thick sheet of PZT, premetallized with 12 µm of fired-on silver, using a wafer-dicing saw with a ceramic-cutting blade. To complete the circuit, the top terminal of the PZT coupon was wirebonded to the PCB.

The devices were then coated with roughly 10 µm of parylene-C for insulation using chemical vapor deposition (Specialty Coating Systems, Indianapolis, IN). No adhesion promoter was used. Parylene was chosen as the insulation material due to its excellent properties as a moisture and chemical barrier[99]. Furthermore, because parylene can be precisely deposited with thicknesses on the order of microns, it does not damp the vibrations of the piezoceramic as strongly as other insulation materials such as polydimethylsiloxane (PDMS) or epoxy, allowing for higher efficiency power harvesting. The stimulation electrodes were exposed by gently scoring the parylene around the electrodes using a tungsten probe tip and peeling the parylene away. The parylene encapsulation used in this work is expected to last from several months to a few years in the biological environment[100]. Implementation of recent advances for long-term conformal encapsulation for mm-scale implantable devices could enable decades- or lifetime-longevity[77,78].

To aid in surgical placement of the mote on the rat sciatic nerve, a cuff was affixed to the back of the mote. The cuff was made by cutting a small piece of silicone tubing (0.8 mm ID, 1.6 mm OD) to roughly the length of the mote. The tubing was then cut in half and attached next to the electrodes using UV-curable epoxy and silicone adhesive.

The volume of the mote stimulator and the combined stimulator and cuff were measured via measuring fluid displacement. To achieve high accuracy, the mass of displaced fluid was measured and volume calculated from the density of the working fluid. This was performed by dropping the device into a container with working fluid on a scale. The object became submersed in the fluid, but was suspended on a wire mesh supported off of the scale. Thus, the buoyant reaction force was measured by the scale, but not the total mass of the object. A single stimulator and cuff (N = 1 of the physical device) were, each separately, measured 7 times (N = 7 technical replicates). The measurement range of volume measurements was reported at 4 times the standard deviation of measurement values observed to take into account additional bias and variance from scale readings, liquid density uncertainty, liquid evaporation, and small volume differences device-to-device. Technical replicates were used to constrain the error of the volume measurement for a very low-volume device.

A further-miniaturized revision of the StimDust mote was also designed (Supplementary Fig. 2). For this design, an IC was fabricated with its size reduced by removing unnecessary test circuits and debug pads and using pad-over-circuit. Additional size reduction in the mote design came from the use of a 0.5 mm x 0.25 mm footprint storage capacitor and refinement of component and wirebond placement. While this smaller design was not used for the experiments reported in this paper, it did achieve a volume of 0.65 mm$^3$ with maximal dimensions of 1.17 mm length, 0.77 mm width, 0.89 mm height (estimated from 3D model). This design is substantially smaller than the tip of a DBS lead. This design also implemented a 'shut-off' timer at 1ms by counting multiple TDC cycles during stimulation. This was a protection against damaging protocols, unintended state programming, or improper start-up. We also implemented a higher current variant of the Stim IC where the reference current generator (Fig. 2c) was scaled by a factor of 5, resulting in output current levels ranging from 250 µA to 2mA with 250µA increments. These current levels are comparable to chronically implanted commercial stimulators which must account for increased stimulation thresholds due to fibrotic or glial scarring, or reduced proximity to the nerve. The high current variant utilized the same mote design featured in Supplementary Fig. 2.

**Design of the external transceiver:** The external transceiver was a compact device designed to be held by a user with its ultrasonic transducer head placed against the skin of a subject. A previous version demonstrated that the external transceiver could also be configured as a wearable device with a low-profile transducer[41]. The transceiver established an ultrasonic wireless link with the mote to provide power, control stimulation parameters via time-coded downlink data, and report back whether the mote had successfully applied each stimulation pulse via detecting backscatter modulation. It was designed around two major components: a high-performance microcontroller (NXP LPC4370) for system control and digital modulation / demodulation, and a custom ultrasound interface IC[101]. Due to the relatively low frequencies used, a full-digital transceiver was implemented with direct digital synthesis of the transmit signal and digitization of the raw receive signal, enabling greater modulation / demodulation design flexibility.

Signal flow through the transceiver is depicted in Fig. 2. A stimulation protocol was programmed into the system and stored in a stimulation table as a sequence of A, W, G, S duration values for each stimulation event. This table could store up to 4500 unique stimulation events of differing parameters and is repeated. An on-board PLL (phase-locked loop) generated a 1.85 MHz square wave carrier signal. On-board timers utilized the stimulation table data to encode the downlink timing with on-off keying onto the carrier signal with sub-microsecond resolution. The ultrasound interface IC used a high-voltage rail provided by an external power supply (5 to 36 V) to level-shift the low voltage transmit signal from the microcontroller to a high-voltage signal used to drive the external piezoelectric ultrasound transducer. Two different ultrasound transducers could be alternately used: one for interfacing with motes at shallow depth (12.7 mm diameter, 21.6 mm focal length, V306-SU-F0.85IN-PTF, Olympus NDT, Waltham MA), and one for interfacing with deep motes (25.4 mm diameter, 47.8 mm focal length, V304-SU-F1.88IN-PTF). For *in vivo* tests using these commercial, metal-encased transducers, a thin sheet of latex was stretched over the transducer face and castor oil was used as coupling between the transducer face and the latex. This prevented electrical paths from the stimulation site to external transceiver ground through the transducer casing. The electromechanical frequency response of the transducer filtered the square wave transmit signal resulting in a sinusoidal acoustic-domain signal. A single-element external transducer was chosen due to the ease and selection of commercial-off-the-shelf options. Depending on the clinical use case, a wearable external transceiver with a multi-element low-profile external transducer, as demonstrated in[41], may enable more convenient long-term stimulation and beamforming for fine-tuning of the beam axis relative to the mote position[62]. In clinical use-cases where the ultrasonic link is to be maintained for long durations, an ultrasound-transmissive pad[102] or adhesive layer[103] could be used in place of ultrasound gel for coupling. Any transmission inefficiencies through this coupling layer could be offset by increased transceiver transmit power.

During ultrasound-free intervals (states A and G), the ultrasound interface chip switched from transmit (Tx) to receive (Rx) mode to pick up the backscattered signal captured by the external transducer. The receive signal was band-pass filtered and amplified (LT6203) by 4.8 dB (adjustable). A 12-bit ADC on the microcontroller digitized the receive signal at 17 MSPS (adjustable). The digitized backscatter was band-pass filtered and broken into two regions based on the timing values of the downlink modulation signal. The first backscatter region, $x_{S,n-1}$, was captured immediately after the transceiver $S_{n-1}$ to $A_n$ transition and had a duration of twice the acoustic time-of-flight between the external transducer and the mote (~ 20 µs to 60 µs depending on mote depth). This corresponded to the duration when the external transducer was no longer driven, but the acoustic wave was still travelling to and reflecting off the mote. The $x_{S,n-1}$ region had an amplitude dependent on the mote modulation state during the mote S period, which corresponded to the mote $M_{mod}$ off. The second backscatter region,

$x_{W,n}$, was captured immediately after the transceiver $W_n$ to $G_n$ transition and also had a duration of two time-of-flight. This region had an amplitude dependent on the mote modulation state during the mote W period. When the mote was properly stimulating, $M_{mod}$ was on during this region, otherwise $M_{mod}$ was off. When turned on, $M_{mod}$ reduced the electrical impedance seen by the mote piezo (Supplementary Fig. 1), reducing the amplitude of the backscattered wave.

To detect the modulation or uplink data for each pulse n ($d_n$), the normalized difference between the time-integrated amplitudes (L1 norms) of the $x_{W,n}$ and $x_{S,n-1}$ backscatter signals were compared to a threshold.

$$d_n = \frac{\|x_{W,n}\|_1 - \|x_{S,n-1}\|_1}{\|x_{S,n-1}\|_1} < thresh$$

The normalization cancelled out any variation due to downlink power or path loss. With calibration, a threshold was determined which separates the backscatter relative difference into "1" bits below the threshold and "0" bits above the threshold. Calibration was performed by gathering backscatter for both "0" and "1" bits, performing a gaussian fit on each population, and finding the threshold which minimized the estimated bit error rate given the population statistics. Either the L1 norm (volt-seconds) or the L2 norm (proportional to backscatter energy in Joules) could be used for this comparison, but L1 showed slightly better separation between backscatter states and was used. In some instances of Rx backscatter capture, there was some clipping of the receive voltage on the high side of the signal due to improperly set gain, but this did not preclude further processing. This single-bit backscattered data from the mote, which signified whether the mote was stimulating or not, was given as feedback to the operator of the handheld external interrogator via an LED when the mote was detected as active. The backscatter bit and raw backscatter digitized waveforms could also be streamed to a PC via a serial connection for further logging and diagnostics, though this was not necessary for normal operation (backscatter data presented in the figures and other results came from this offload connection).

Custom C code was used to program the external transceiver microcontroller. LPCOpen (v2014.05) and CMSIS_DSPLIB (v2013) C code libraries from NXP (NXP Semiconductors N.V.) were utilized in creating this code. ARM gcc (arm-none-eabi-gcc GNU Tools for ARM Embedded Processors 6-2017-q2-update 6.3.1 20170620 ARM/embedded-6-branch revision 249437) and MCUXpresso IDE (10.1.1.606, NXP) were used in development, compiling and deployment of transceiver code. The external transceiver collected backscatter data and transferred it to a Matlab script (R2017b, R2019a Mathworks, Inc) running on a PC which coordinated experiments and saved data.

**Characterization of motes in ultrasound gel tissue phantom and in *ex vivo* tissue:** To assess performance of the system, a mote with test leads was placed in ultrasound gel (Aquasonic Clear, Parker Labs, Fairfield NJ), either freely-floating or mounted on top of a 700 μm diameter steel rod connected to a rotation stage (RP01, Thorlabs, Newton, NJ). Bubbles were removed from all ultrasound gel used in this study via centrifugation for 16 minutes at 428 x $g_0$ to 1052 x $g_0$. Ultrasound gel has acoustic impedance and attenuation within 0.7% and 1.9%, respectively, of those of water[104]. An electrical load was attached to the mote electrode terminals to mimic tissue electrical properties (series 3 kΩ and 22 nF were used; note that in-vivo values were $R_{Soln}$ = ~4.54 kΩ and $C_{dl}$ = ~100 nF). The external transceiver was mounted to a computer-controlled 2-axis stage (XSlide, VelMex) with the external transducer contacting the gel. Mote parameter encoding and decoding performance were assessed at a mote depth of 18 mm using the 12.7 mm diameter external transducer at a transmit voltage of 19.4 $V_{pk-pk}$. Stimulation, power and backscatter performance were assessed at a mote depth of 48 mm using the 25.4 mm diameter external transducer at a transmit voltage of 25.6 $V_{pk-pk}$. Backscatter modulation uplink performance was assessed by determining the separation of demodulated 'one' bits and 'zero' bits. 'One' bits were gathered with a fully-powered mote showing a high relative difference between the modulation states ($M_{mod}$ off during $S_{n-1}$ and $M_{mod}$ on during $W_n$). 'Zero' bits were gathered by operating a mote slightly underpowered so that stimulation did not occur and there was a low relative difference between the modulation states ($M_{mod}$ off during $S_{n-1}$ and $M_{mod}$ off during $W_n$). Each backscatter experiment consisted of 400 or more pulses. The first and last up to 20 pulses of data in each backscatter experiment were thrown out due to interfering experimenter activity in the vicinity of the experiment during initiation and stoppage of the experiment, which caused increased noise (the first and last 2 for *in vivo* data in Fig 6e). The minimum acoustic intensity needed to operate StimDust was measured on a selection (N=6) of the fabricated devices which passed initial acceptance testing. The final value was calculated as the average of the best-performing one-third (N=2) of these devices. The performance of the system at various positions and alignments of the mote in the acoustic link were assessed by rastering over the transverse translational and angular offset dimensions at a depth of 48 mm with the 25.4 mm diameter external transducer. The performance of the system at various depths and at high pulse repetition frequency were assessed with the mote on the central axis of the external transducer using the 25.4 mm diameter external transducer at a transmit voltage of 31.5 $V_{pk-pk}$. The external transceiver acoustic field for each transducer was characterized in a separate experiment in a water tank using a hydrophone (HGL-0400, Onda, Sunnyvale CA USA) (Fig S3). The system power transfer efficiency values at various points in energy transfer between the external transceiver to the acoustic medium to the mote storage capacitor to the charge delivered to the tissue was calculated using: voltage and current draw of the external transceiver ultrasound driver; hydrophone-obtained acoustic field data integrated over 1 cm radius circular area in the focal plane; hydrophone-obtained acoustic field data integrated over the face of the mote piezo; voltage charge waveform of the mote storage capacitor acquired via oscilloscope; calculated energy and power waveforms into the energy storage capacitor given its capacitance; and calculated stimulation pulse charge, energy, and power from measured stimulation voltage, stimulation current, and stimulation duration. Amplitudes were determined via IQ demodulation envelope detection. Calculated power-harvest traces were smoothed.

StimDust performance at depth in tissue was assessed by operating the device through fresh *ex vivo* porcine tissue. The posterior compartment of a pig hind leg was positioned between the external transducer and the mote, with coupling provided by ultrasound gel. The ultrasonic wireless link passed through approximately 1.5 mm of skin, 1 mm of fat, 47.5 mm of musculature, including the gastrocnemius and soleus muscles, and 5 mm of gel.

***In vivo* extracellular electrophysiological stimulation:** All *in vivo* procedures were performed in accordance with University of California-Berkeley Animal Care and Use Committee regulations. Wireless stimulation was performed in six animals. Stimulation was confirmed in all animals via visual observation of the muscle twitch. Detailed quantitative results were collected for animals C, D, and F. For each procedure, a male Long-Evans rat (Charles River, Wilmington, MA) of approximately 300 g was anesthetized with 1250 mg/kg ethyl carbamate and placed on an isoflurane vaporizer at 0.5% isoflurane and 300 mL/minute flow rate. The skin around the leg was shaved and depilated. A semi-circular flap incision was made from approximately the midpoint of the tibia to the midpoint of the femur, following the curvature of the knee joint. The biceps femoris was longitudinally bisected and resected to uncover the sciatic nerve. The fascia holding the sciatic nerve to the underlying musculature was removed, but the epineurium was left intact. The area was irrigated with sterile saline to ensure good electrical contact between the mote and the nerve. A mote was placed on the nerve approximately 4 mm proximal to the bifurcation of the nerve to innervate the limb extremities (Fig. 1b). While this surgery was not optimized for minimal invasiveness, further practice and refinement of the surgical procedure and technique, in addition to development of tailored laparoscopic tools, would enable implantation of the mote with a vastly smaller incision. The mote had 5 test leads attached to it for data collection purposes (GND, $V_{Pz+}$, $V_{Pz-}$, $V_{DD}$, $V_{elec}$) (Fig. 1c). These leads were unnecessary for the function of the mote and were used for debugging, recording wired data to compare with the wireless readout for characterization purposes, and to standardize the fine-alignment of the acoustic link; they provided no power or communication signals to the mote. To ensure that these leads had no effect on the performance of the system, they were completely disconnected from all external instrumentation during a test trial and no change in system performance was seen. Figures 1b, 1c, S12b, and S12c were acquired with a Summit K2 USB microscope camera (OptixCam). Other images were taken with a modern consumer smartphone camera. Images were lightly edited for crop, brightness (uniformly applied), and contrast (uniformly applied).

For initial verification of functional stimulation and sweeps of stimulation current and pulse width, the surgical site was left open and the external interrogator was mounted such that the external transducer face was approximately 20 mm from the mote, with ultrasound gel filling the intervening space. The external interrogator was then turned on and chip $V_{DD}$ was measured from test leads on the mote. The external interrogator position was adjusted to ensure good fine-alignment with the mote and successful stimulation was verified with test leads on the mote. This use of a measure of $V_{DD}$ for fine alignment of the mote was used for repeatability between trials but was not necessary for function of the system. To verify this, positioning of the external transducer relative to the mote was performed without using $V_{DD}$ as an alignment reference and the mote was able to receive enough power to function. The backscatter amplitude was also an option to aid in positioning the mote without use of a $V_{DD}$ measurement.

Two penetrating stainless-steel electrodes were placed in the gastrocnemius muscle to capture the electromyogram signal (EMG) induced by sciatic nerve stimulation. The EMG signal was amplified by 100X to 1000X, bandpass filtered at 10 Hz to 3 kHz (DP-304, Warner Instr. Co., Hamden CT), digitized at 100 kSPS (NI DAQ USB 6001, NI, Austin TX), and collected with Signal Express (v2015, National Instruments) software. In one animal (Animal F), EMG recordings were taken from three downstream muscles: the medial head of the gastrocnemius, the lateral head of the gastrocnemius, and the tibialis cranialis (tibialis anterior in bipeds).

The external interrogator was programmed to specify a stimulation current, pulse width, and pulse repetition frequency. The system was initiated and performed several to tens of stimulation events per condition before being set to different stimulation parameters. The pulse repetition frequency was kept at a relatively low ~1/3 Hz while generating recruitment curves to prevent muscle fatigue. Stimulation of the sciatic nerve elicited compound muscle action potentials (CMAP) across much of the lower leg musculature. CMAP amplitudes were calculated from baseline to the first peak (excluding the stimulation artefact). To calibrate the stimulation response curve, mote parameters were swept across stimulation current and stimulation pulse width dimensions, and baseline-to-peak CMAP amplitudes were calculated to construct a recruitment curve. A sweep of pulse repetition frequency was also performed and qualitative muscle response recorded. A rare ultrasound downlink decoding error was observed in which the mote watchdog timer incorrectly identified a change of ultrasound state, resulting in a stimulation of different duration than that specified. These stim pulses were excluded from the recruitment curve analysis (1 total with animal C, 0 total with animal D, 4 total with animal F). An erroneously long stimulation pulse from one of these decoding errors resulted in tissue discoloration. EMG traces presented in the figures and used for calculating amplitudes were digitally band-pass filtered from 50 Hz to 2,550 Hz after acquisition. Mote electrode and $V_{DD}$ recorded voltages presented in figures were digitally low-pass filtered at 2.4 MHz. Mote piezo recorded voltages presented in figures were digitally low-pass filtered at 8.4 MHz. Data was analysed with custom Matlab scripts (R2017b and R2019a).

To verify that the acoustic link was functional while transmitting through the tissue layer of the animal, the experiment was repeated with a closed surgical site (Fig 1d). Excess ultrasound gel was removed, the biceps femoris replaced over the nerve and mote, and the skin incision sutured with 6-0 nylon suture. With the wound closed, ultrasound gel was applied to the outside of the animal and the external interrogator was repositioned with the external transducer in contact with gel so that the distance from transducer to mote was 18 mm, with approximately 5 mm consisting of tissue. Downlink, stimulation, and uplink performance were then assessed with the fully implanted mote.

To characterize *in vivo* telemetry, backscatter modulation was detected as described above. The relative difference between the L1 norms of these two regions was calculated for 111 successful *in vivo* stimulation pulses (Fig. 6e left population). The relative difference between the L1 norms of the $S_n$ and $S_{n+1}$ backscatter regions was calculated for the same 111 pulses (Fig. 6e right population) and was used as an estimate of the backscatter when $M_{mod}$ does not turn on during the W phase (which is the case when the mote does not stimulate).

The *in vivo* data presented in Figures 6, 7, S5 and S6 originate from 18 sets of trials with animal C, 51 sets of trials with animal D, and 21 sets of trials with animal F. An additional 52 sets of trials with animal C, 102 sets of trials with animal D, and 152 sets of trials from animal F were performed. For all *in vivo* studies, many trials were performed to ensure that the system was set up and operating properly, to generate video data depicted in video S1, and for final 'exploratory' sets of trials taken after the data shown in the figures. All data for every animal and every trial is available upon reasonable request.

**Data Availability:** Data used to support results and draw conclusions are presented in this paper and associated Supplementary Information. Figures 4, 5, 6, 7, S3, S4, S5, S6, S7, S8, and S9 have associated raw, pre-processed data. This raw, pre-processed data is available and is linked at https://doi.org/10.6084/m9.figshare.11719611.

**Computer Code Availability**: Computer code used for analysing data is available at https://github.com/maharbizgroup/StimDust.

**Acknowledgements**

This work was supported in part by the National Institutes of Health NIH R21EY027570, The Defense Advanced Research Projects Agency (DARPA) BTO Grant/Contract No. HR011-15-2-0006 under the auspices of Dr. Doug Weber, The National Science Foundation NSF EAGER 1551239, The McNight Foundation Technological Innovations in Neuroscience Award [M.M.M., J.M.C.], The Chan Zuckerberg Biohub [R.M., M.M.M.], NIH Training Grant T32 GM008155.

**Author Contributions**

B.C.J., R.M., M.M.G. and K.L. designed and characterized the integrated circuit.

D.K.P., J.E.K. and M.M.M. designed and implemented the external transceiver.

K.S., D.K.P., M.M.M. and R.M.N. designed and assembled the motes.

D.K.P., R.M.N., J.M.C., K.S. and B.C.J. designed and carried out the *in vivo* studies.

D.K.P. performed *in vitro* and *ex vivo* studies.

D.K.P. and B.C.J. conducted data analysis and simulations.

All authors discussed results.

D.K.P., B.C.J., R.M., M.M.M., J.M.C., K.S. and M.M.G. prepared the manuscript with input from all authors.

**Competing Interests:** The authors declare the following competing interests: M.M.M., J.M.C., R.M.N, and J.E.K. are members of iota Biosciences, Inc.

# Supplementary Information for:

# StimDust: A mm-scale implantable wireless precision neural stimulator with ultrasonic power and communication

David K. Piech[1]*, Benjamin C. Johnson[2,3]*, Konlin Shen[1], M. Meraj Ghanbari[2], Ka Yiu Li[2], Ryan M. Neely[4], Joshua E. Kay[2], Jose M. Carmena[2,4]**, Michel M. Maharbiz[2,5]**, Rikky Muller[2,5]**

[1] University of California, Berkeley – University of California, San Francisco Graduate Program in Bioengineering, University of California, Berkeley, Berkeley, CA USA 94720

[2] Department of Electrical Engineering and Computer Sciences, University of California, Berkeley, Berkeley, CA USA 94720

[3] Department of Electrical Engineering and Computer Engineering, Boise State University, Boise, ID USA 83725

[4] Helen Wills Neuroscience Institute, University of California, Berkeley, Berkeley, CA USA 94720

[5] Chan-Zuckerberg Biohub, San Francisco, CA USA 94158

* Co-first authors: D.K.P., B.C.J.;    ** Co-senior authors: R.M., M.M.M., J.M.C.

** e-mail: R.M. (rikky@berkeley.edu); M.M.M. (maharbiz@berkeley.edu); J.M.C. (jcarmena@berkeley.edu)

# Contents

# Supplementary Section 1: Additional results from characterizing mote performance

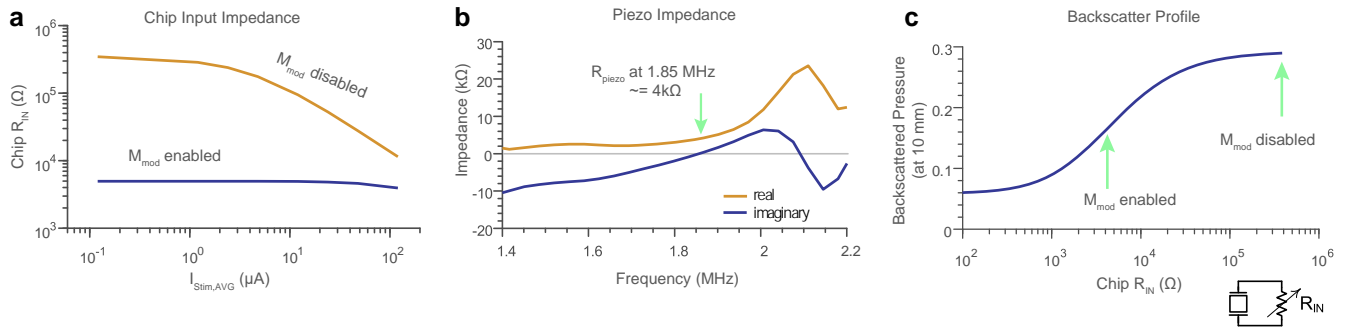

**Figure S1 | Backscatter modulation design. a**, The measured chip input resistance ($R_{IN}$) for different current stimulation loads. The bottom trace is when the modulation switch is enabled, keeping $R_{IN}$ constant across load. The difference between the two lines is the modulation depth for a given load. **b**, Simulated mote piezo impedance versus frequency. At resonance, the piezo has a 4 kΩ impedance. **c**, Simulated backscatter profile at 10 mm based for different $R_{IN}$. N = 1 sweep for the data presented in panel (a).

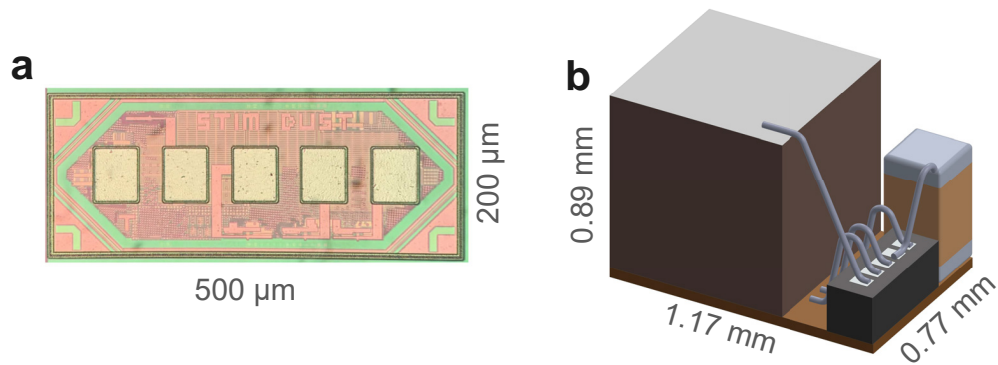

**Figure S2 | Further miniaturized StimDust mote. a,** fabricated further-miniaturized StimDust IC (integrated circuit). **b,** Design of further-miniaturized mote utilizing the fabricated IC and a piezo of the same dimensions as that in Fig 4.

The step response ring-up time of the external transducer was approximately 1 cycle, or 0.56 μs. The ring-up time for the external transducer and mote piezo system (full downward link) was approximately 3.5 cycles or 1.94 μs. The ring-up time for the external transducer, mote piezo, and backscatter capture system (full bidirectional link) was approximately 4.5 cycles or 2.5 μs. A pulse-train mode was demonstrated by interspersing several pulses of short-duration shorting-phase with a pulse of long-duration shorting-phase. This yielded a pulse-train every 500 ms, with each train containing 10 pulses of 100 μs pulse width occurring every 750 μs.

The acoustic beam patterns of the 12.7 mm and 25.4 mm diameter external transducers (Supplementary Fig. 3) were had a focal plane intensity transverse full-width-half-maximum of 1.6 mm and 1.8 mm, respectively, and an axial intensity longitudinal full-width-half-maximum of 14 mm and 17.7 mm.

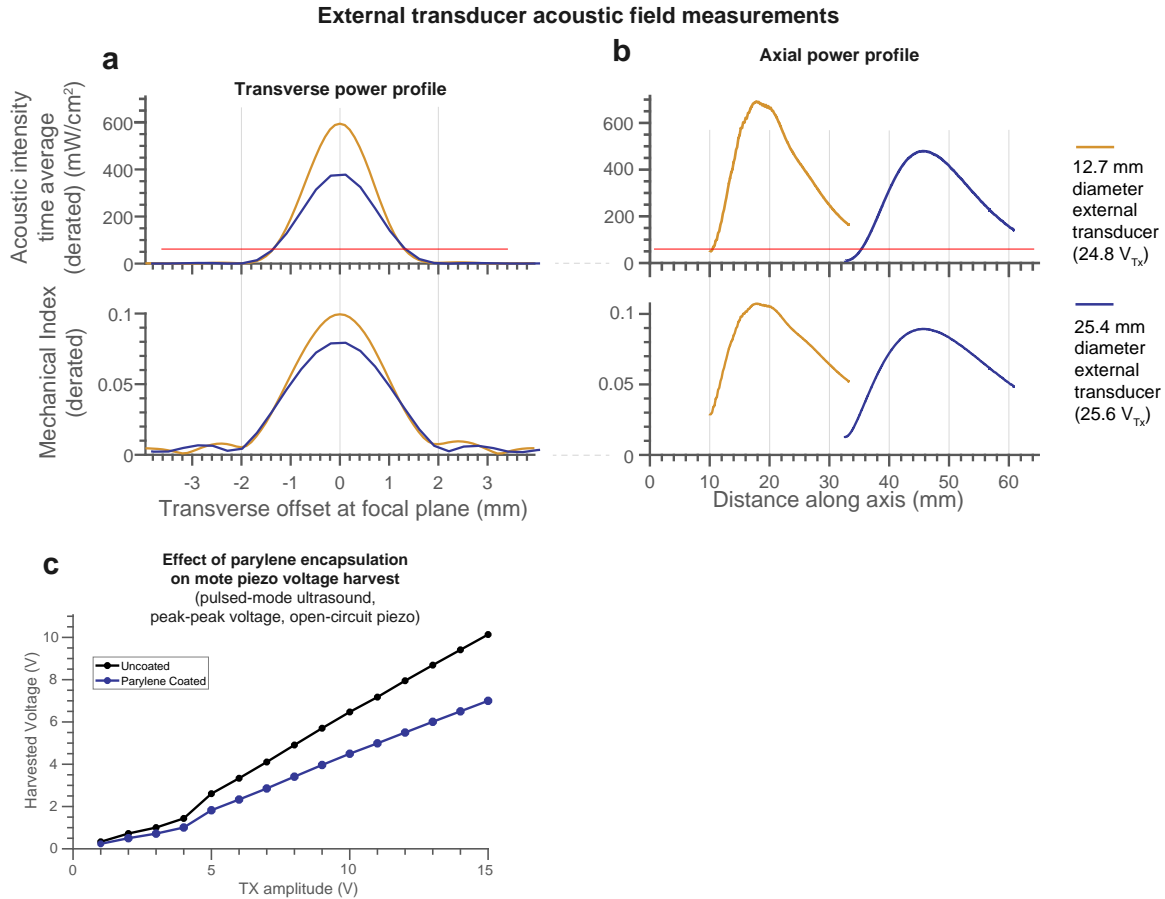

**Figure S3 | External transceiver acoustic field and piezo voltage harvest.** Characterization of the external transceiver acoustic field for two alternate transducers used for shallow (12.7 mm diameter transducer) and deep operation (25.4 mm diameter transducer). **a**, transverse beampattern. **b**, longitudinal beampattern. Data taken in a water tank with external transceiver and hydrophone. Red lines indicate the minimum acoustic intensity necessary to operate StimDust. **c**, Effect of parylene encapsulation on mote piezo voltage harvest. (a) Shown here is N = 1 fine-stepped sweep of transverse offset; a different coarse-stepped sweep of transverse offset produced similar results. (b) Shown here is N = 1 fine-stepped sweep of axial offset; a different coarse-stepped sweep of axial offset produced similar results. (c) N = 1 voltage sweep for each condition.

The power performance of the system was measured for two conditions: benchtop operation at high PRF (pulse repetition frequency) and *in vivo* operation with a fully implanted mote at low PRF (Supplementary Table 1). $V_{TX}$ and $P_{electrical}$ into the external transducer were measured at the output of the external power supply used to supply the ultrasound interface chip. The external transducer was heavily damped to shorten its impulse response and much of the acoustic power generated in the external transducer piezo was dumped into the absorptive backing layer, causing the electrical input power to acoustic power at the focal plane efficiency to be low.

Acoustic domain measurements were made with a hydrophone in water; notably, the hydrophone measurements were taken in a separate experiment from mote operation and the acoustic values in table 1 assume that the mote was positioned at the point of maximal acoustic intensity (this may cause up to ~20% error). For the *in vivo* condition, the hydrophone measurements could not be made directly, and so expected loss from impedance mismatches and absorption in 2 mm of skin and 3 mm of muscle ([1], [2], [3]) was modelled to yield an estimate of acoustic intensity inside the animal. The acoustic power at the focal plane (depth of the mote) was integrated over a 1 cm radius circle, which is at least 97% of the total power in the focal plane. The face of the mote captured approximately 20% of the acoustic power in the focal plane. This was a trade-off that balances minimizing the power coupling sensitivity of small transverse misalignments with minimizing the unused ultrasound power radiating into the body.

The acoustic power conversion efficiency ($\eta_{acoustic}$) is defined as the ratio of the electrical power used to charge to the mote relative to the acoustic power at the face of the mote piezo. This efficiency encompasses the acoustic to electrical power efficiency of the piezo, the rectifier efficiency, and the standby power consumption of the IC. It was calculated from the measured rate of change of the $C_{store}$ capacitor voltage ($V_{DD}$) during initial mote power-up for a 3V $V_{DD}$ steady-state. Under a given incident acoustic power, the $\eta_{acoustic}$ (as measured by P into $C_{store}$) peaked at ~8.1% when $V_{DD}$ was near half the steady-state voltage (Fig. 6f, Supplementary Fig. 4). The mote's stimulation power ($P_{stimulation\_delivered}$) was calculated by measuring the voltage across an electrode model load. The electrode load was modelled as a solution resistance ($R_{soln}$) and double-layer capacitance ($C_{dl}$) in series. $R_{soln}$ and $C_{dl}$ were discrete passives for the benchtop test; *in vivo* values were estimated from the stimulation voltage at known current output by using the initial IR drop (current times resistance voltage drop) and initial slope of capacitive charging. $Q_{pulse}$ (charge per pulse), $E_{pulse}$, and $P_{stimulation\_delivered}$ were calculated from the measured stimulation voltage waveform, $R_{soln}$, $C_{dl}$, and $f_{stim}$. $P_{stimulation\_available}$ was calculated from $Q_{pulse}$ delivered, the voltage headroom available, and $f_{stim}$. $P_{stimulation\_delivered}$ represents the power delivered to the electrode model used (and would change with load) while $P_{stimulation\_available}$ represents the maximum stimulation power available for any load.

Note that at low PRF the total $P_{stimulation\_delivered}$ was low and the device charged up quickly and remained 'idle' for most of the time between stimulation pulses yielding a low overall efficiency. Under low PRF conditions, it was shown that the device could be intermittently powered up for each pulse to reduce the overall ultrasound dose. If optimized for this application, the mote could have a small $C_{store}$ to reduce start-up time.

**Table S1: System power performance**

| | Benchtop high PRF example | *In vivo* low PRF example |
|---|---|---|
| Acoustic medium | ultrasound gel | gel, skin, muscle |
| Mote depth from ext. txdr | 48 mm | 18 mm |
| **Stimulation protocol** | | |
| Specified PRF | 2380 Hz | .222 Hz |
| Specified stim current | 400 μA | 400 μA |
| Specified stim pulse width | 72 μs | 172 μs |
| Specified interphase gap | 10 μs | 80 μs |
| **External transceiver** | | |
| External transducer | 25.4 mm Ø | 12.7 mm Ø |
| $V_{TX}$ | 28.9 V | 24.8 V |
| U/S duty cycle with UFI's | 76% | 100% |
| $P_{electrical}$ drive ext. txdr | 2.0 W | 1.34 W |
| $I_{SPTA}$ derated | 723 mW/cm$^2$ | 713 mW/cm$^2$ |
| $P_{acoustic}$ at surface over txdr face | 20.9 mW | 17.7 mW |
| $P_{acoustic}$ at focal plane in 1 cm radius | 20.8 mW | 14.6 mW |
| $P_{acoustic}$ on mote piezo face | 4.3 mW | 2.8 mW |
| **Mote** | | |
| Mote $C_{store}$ | 4 μF | 4 μF |
| $V_{DD}$ steady state | 1.9 V | 3.0 V |
| P into $C_{store}$ average | 147 μW* | 9.7 μW |
| $\eta_{acoustic\_mote\_Face \rightarrow electrical\_VDD}$ | 3.4%* | 0.34% |
| $\eta_{acoustic\_at\_surface \rightarrow electrical\_VDD}$ | 0.70% | 0.06% |
| **Load** | | |
| Load $R_{soln}$ (est. for *in vivo*) | 3 kΩ | 4.4 kΩ |
| Load $C_{dl}$ (est. for *in vivo*) | 22 nF | 100 nF |
| $Q_{pulse}$ | 24 nC | 43.4 nC |
| $E_{pulse}$ | 37 nJ | 108 nJ |
| $P_{stimulation\_delivered}$ | 89 μW | 24 nW |
| $P_{stimulation\_available}$ | 103 μW | 30 nW |

\* these power values were derived based on the incident acoustic power at the mote face and the measured efficiencies at various $V_{DD}$s. ('Ø' is diameter; 'UFI' is ultrasound-free interval; 'ext. txdr' is external transducer; 'rad.' is radius; 'soln' is solution.)

# Supplementary Section 2: Supplementary *in vivo* results

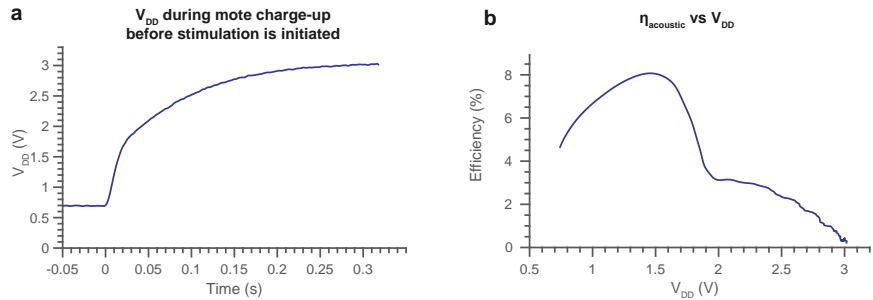

**Figure S4 | *In vivo* power harvesting performance during initial power-up. a,** After 4.2 s without receiving power, the mote $V_{DD}$ (V across $C_{store}$) was ~0.7 V. When downlink power resumed (at t = 0), the 4 µF $C_{store}$ charged over ~300ms. **b,** The charging waveform from (a) was used to estimate the efficiency during charging as VDD passed through various regimes. $\eta_{acoustic}$ is the ratio of electrical power used for charging the mote to acoustic power at the face of the mote piezo. At 1.2 $V_{DD}$, the rate of electrical power harvest was maximum but $V_{DD}$ was below the POR (power-on reset) cut-off and the device did not stimulate. Between 1.9 $V_{DD}$ and 3.0 $V_{DD}$, the mote was able to stimulate. Power harvest efficiency was moderate at 1.9 $V_{DD}$ and decreased to very low values at 3.0 $V_{DD}$, which was the point where $C_{store}$ was nearly topped off and 'trickle charging' under the given incident acoustic power conditions. Representative example out of N = 2 charging periods under this experimental condition.

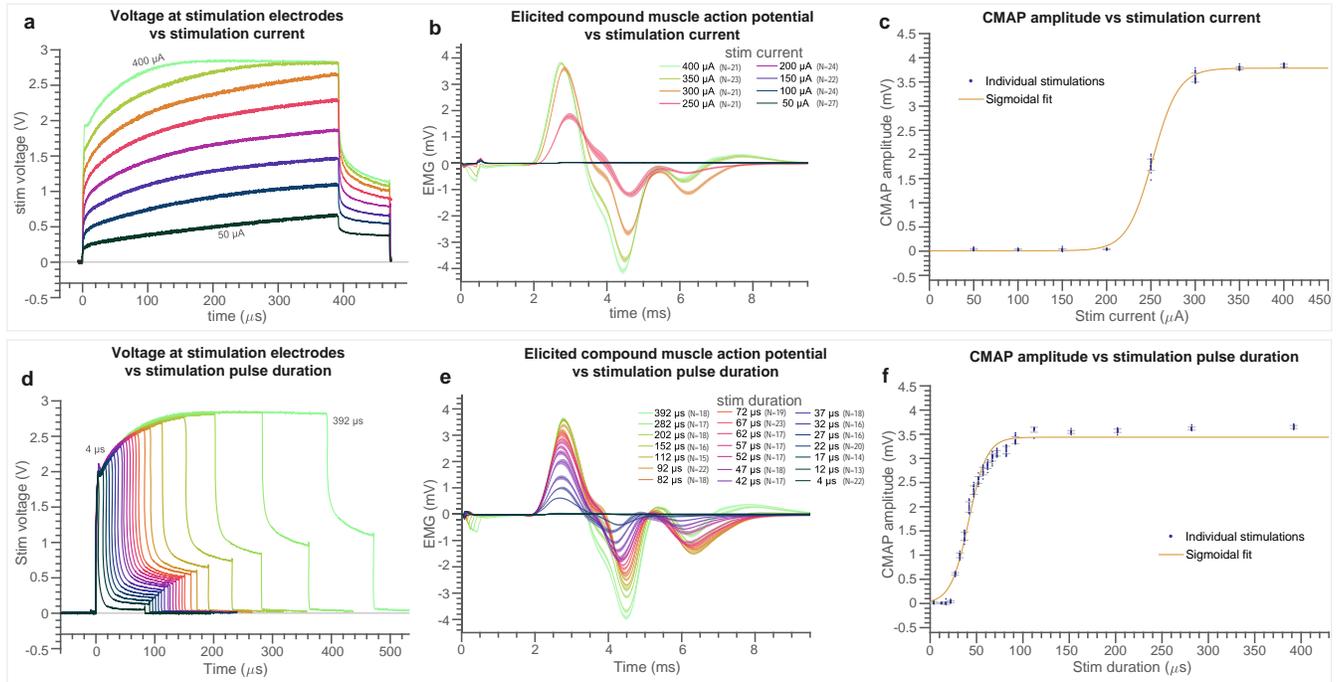

**Figure S5 | Precise control of evoked neural response achieved through varying stimulation current or stimulation pulse duration; second sweep of parameters.** This data was taken with the same animal and mote as that in Fig. 7, but 2 hours after the first sweep. **[Current-control]**: Stimulation current was varied from 50 μA to 400 μA with stimulation pulse width held at 392 μs. **a**, Stimulation electrode voltage shows an approximately linear trend with stimulation current (N = 1 stimulation pulse waveform collected for each condition). **b**, CMAP (compound muscle action potential) waveforms show activation for 250 μA to 400 μA. For each condition, the coloured line represents the trial-average CMAP voltage at each point in time, the width of the shaded error region is ± 1 s.d., and N is the number of pulse events in each condition. The pulse repetition rate of approximately 1/3 Hz yielded essentially independent biological responses to each pulse, though there could have been small non-independent effects due to muscle fatigue. **c**, CMAP amplitude vs. stim current shows a typical sigmoidal recruitment curve. Each point is a single CMAP baseline-to-peak amplitude, the line is a sigmoidal fit and error bars are ± 1 s.d. of the CMAPs in each stim current condition. The sample size for each condition in panel c is the same as that for each condition in panel b. **[Pulse-width-control]**: Stimulation pulse width was varied from 4 μs to 392 μs with stimulation current held at 400 μA. **d**, Stimulation electrode voltage shows an approximately linear trend with stimulation current (N = 1 stimulation pulse waveform collected for each condition). **e**, CMAP waveforms show activation from 27 μs to 392 μs. For each condition, the line represents the trial-average CMAP voltage at each point in time, and the width of the shaded error region is ± 1 s.d., and N is the number of pulse events in each condition. This panel shows that the duration of the electrical artefact increased with increasing stimulation pulse width, but the time-course of the CMAP did not appreciably change, with only amplitude differing. **f**, CMAP amplitude vs. stim pulse width shows a typical sigmoidal recruitment curve with little or no evoked response below 27 μs pulse width and saturation at approximately 112 μs pulse width. Each point is a single CMAP baseline-to-peak amplitude, the line is a sigmoidal fit and error bars are ± 1 s.d. of the CMAPs in each stim duration condition. The sample size for each condition in panel f is the same as that for each condition in panel e. Note: This data was taken with an open surgical site. All replications in this figure are separate stimulation events in a single animal.

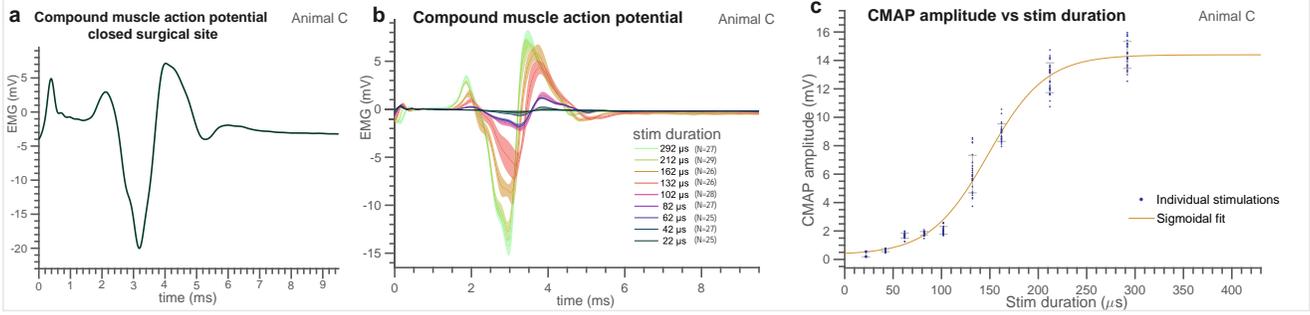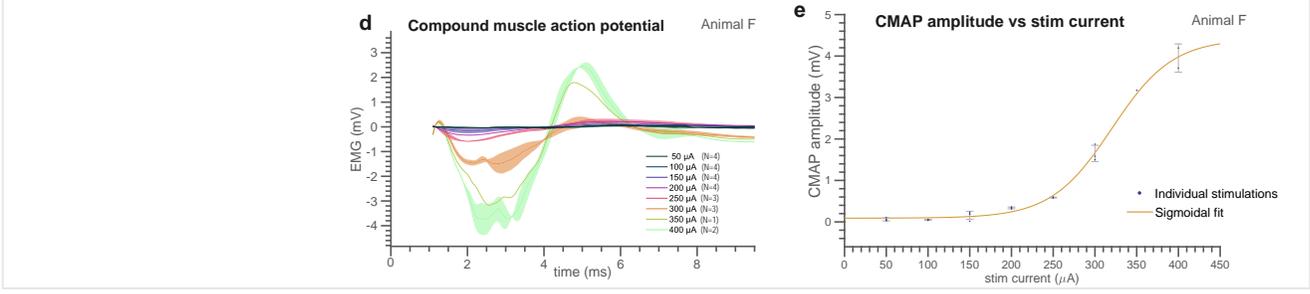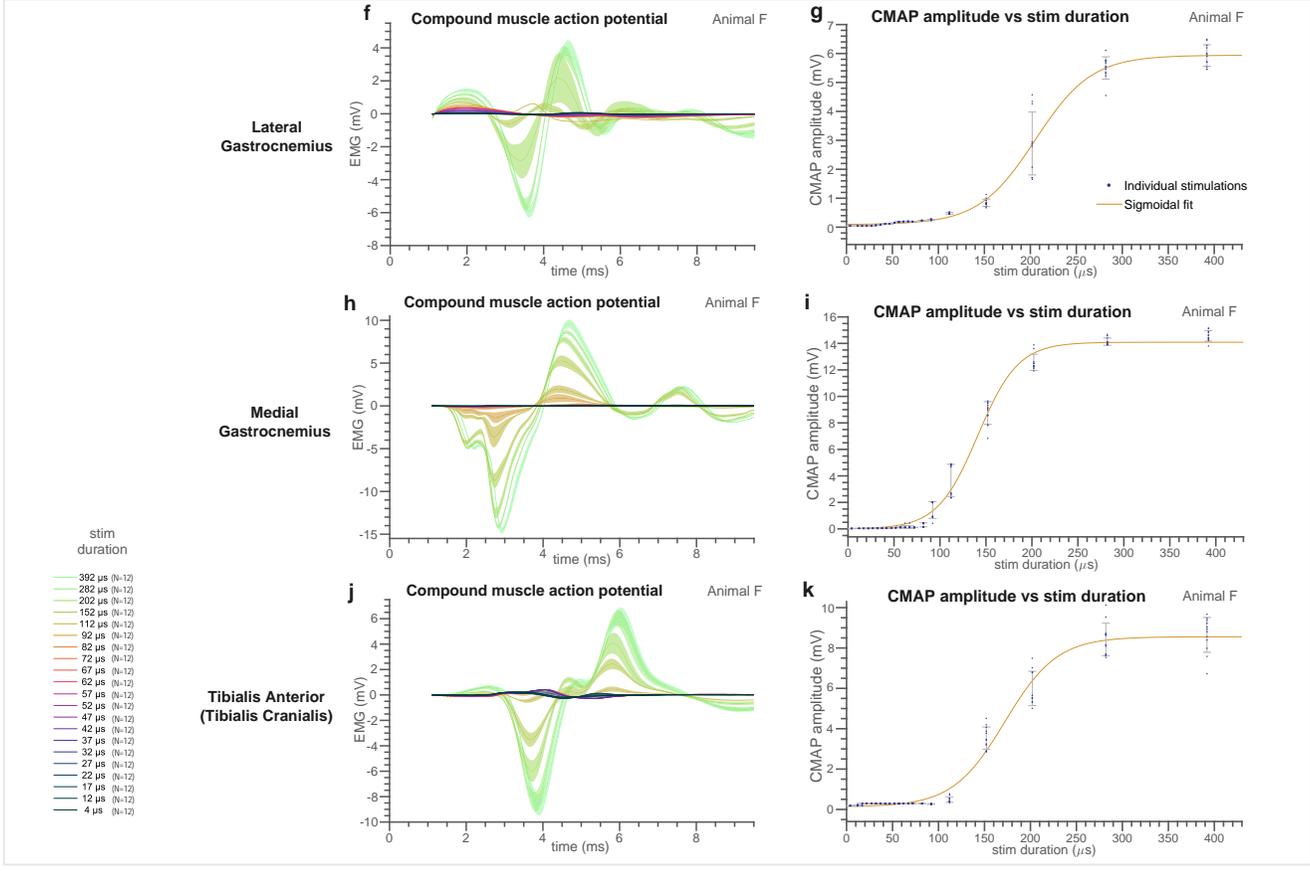

**Figure S6 | Supplementary *in vivo* results: Animals C and F. a**, EMG (electromyogram) trace of CMAP (compound muscle action potential) resulting from stimulation that was performed in Animal C with the ultrasonic link traversing muscle and skin layers between the implantation site and the exterior of the body. Representative example CMAP trace for N = 1 stimulation pulse within an experimental condition with 20 stimulation pulses. **b, c,** StimDust operation in animal C at stimulation pulse duration values between 22 μs and 292 μs, with sigmoidal CMAP recruitment. **d, e**, StimDust operation in animal F at stimulation current values between 50 μA and 400 μA, with sigmoidal CMAP recruitment. **f-k**, StimDust operation in animal F at stimulation pulse duration values between 22 μs and 392 μs, with sigmoidal CMAP recruitment across three innervated muscles. In panels b, d, f, h, and j, for each condition, the coloured line represents the trial-average CMAP voltage at each point in time, the width of the shaded error region is ± 1 s.d., and N is the number of pulse events in each condition. In panels c, e, g, i, and k, each point is a single CMAP baseline-to-peak amplitude, the line is a sigmoidal fit and error bars are ± 1 s.d. of the CMAPs in each stim current condition. The sample size for each condition in panel c is the same as that for each condition in panel b. The sample size for each condition in panel e is the same as that for each condition in panel d. The sample size for each condition in panels f, g, h, i, j, and k is given on the left-side figure legend. All replications of a given stimulation condition in panels **a-e** are separate stimulation events in animal C. All replications of a given stimulation condition in panels **f-k** are separate stimulation events in animal F. The pulse repetition rate of approximately 1/5 Hz for all panels in this figure yielded essentially independent biological responses to each pulse, though there could have been small non-independent effects due to muscle fatigue.

**Video S1 | *In vivo* neural stimulation with fully implanted wireless StimDust.** Mote implanted on rat sciatic nerve with stimulation pulses delivered at 0.22 Hz. In clips (a) and (b) (animal D), the animal's right hindquarters are visible with cranial to the right and caudal to the left. The closed surgical site is at the centre of the frame and the animal's right leg and foot are towards the bottom left of the frame, with two EMG electrodes visible. The external transducer can be seen at the top of the frame. Ultrasound gel fills the gap between the external transducer and the animal, and a glass slide makes this space visible. Clip (c) shows a view from the rear (animal D). Clip (d) shows a different experiment with the mote implanted on the left sciatic (animal C).

# Supplementary Section 3: Discussion supporting the claim that stimulation was due to the mote output current and was not directly ultrasound-mediated

At 1.85 MHz, the threshold of ultrasound-mediated neural stimulation has been reported at approximately 10 W/cm$^2$ pulse-average intensity[4]; this is more than 14x higher intensity than that used in this study. Furthermore, as the system increased stimulation current from 50 μA to 400 μA with a corresponding recruitment of CMAP response, acoustic power was nearly identical and actually decreased slightly since the protocol utilizes a longer TDC (time-delay control) gap when specifying high current. Additionally, no EMG response was observed when the system was driven with continuous ultrasound at the same intensity as used for controlling the device, but with no coded downlink signals. Finally, a pilot *in vivo* experiment, which powered a mote stimulator IC electrically with no ultrasound, produced stimulation and evoked CMAP's similar to those evoked with an acoustically-powered mote (Fig S7). Utilizing a current-control implantable stimulator provides improved stimulation precision and spatial resolution as compared to directly mediated ultrasonic stimulation, and requires substantially lower acoustic intensities and thus has lower risk from ultrasound-induced thermal or cavitation damage[5].

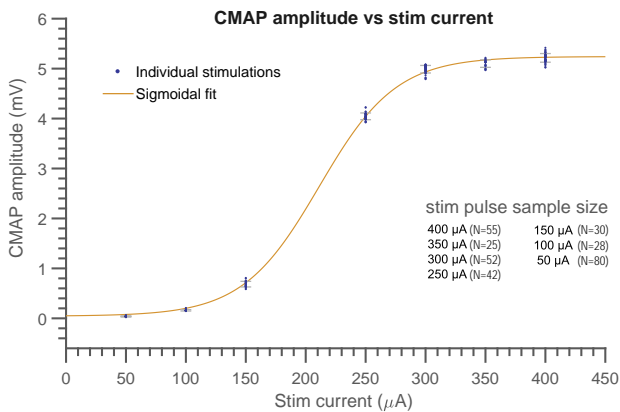

**Figure S7 | Pilot study: *in vivo* stimulation and elicitation of compound muscle action potential with electrically-powered mote.** A cuff was implanted on the sciatic nerve of a rat. The cuff electrodes were attached to a StimDust mote that was assembled on a larger printed circuit board located outside the animal. The electrical terminals of the external transceiver piezo drive circuit were directly connected to the PZ+ and PZ- terminals of the mote. This entirely bypassed the acoustic wireless link (no ultrasound energy was involved), but otherwise utilized all mote functionality and downlink communication protocol. CMAP amplitude vs. stim current shows a typical sigmoidal recruitment curve which is similar to that measured when utilizing the ultrasound wireless link. Each point is a single CMAP baseline-to-peak amplitude, the line is a sigmoidal fit and error bars are ± 1 s.d. of the CMAPs in each stim current condition. N is the number of pulse events in each condition. The pulse repetition rate of approximately 1 Hz yielded essentially independent biological responses to each pulse, though there could have been small non-independent effects due to muscle fatigue. Note: This data was taken with an open surgical site. All replications in this figure are separate stimulation events in a single animal.

# Supplementary Section 4: Discussion of electrodes and monophasic stimulation

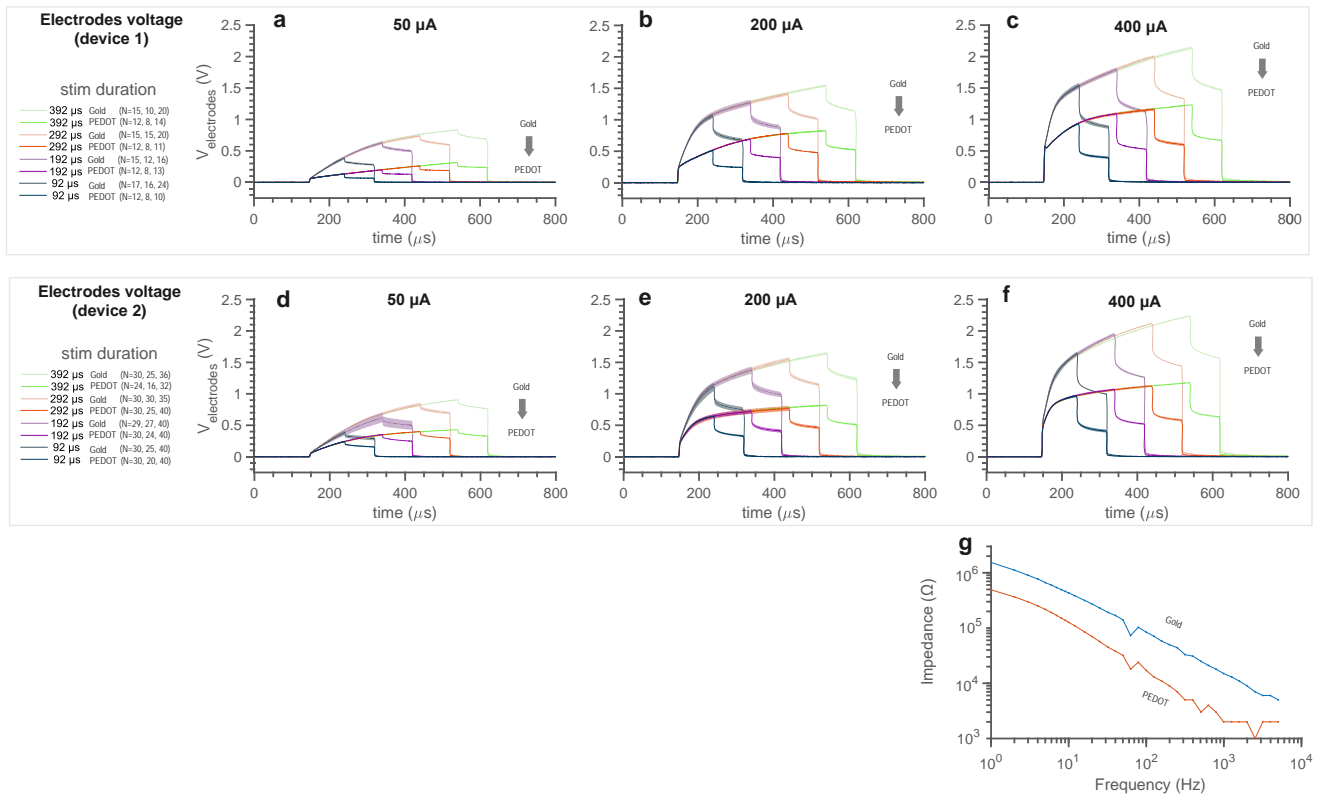

**Figure S8 | Stimulation electrodes characterization in saline. a, b, c, d, e, f**, The electrode voltage during stimulation pulses was measured for two devices before and after PEDOT:PSS (poly(3,4-ethylenedioxythiophene) polystyrene sulfonate) plating (a third device was tested but is not shown due to a suspected short). Each device that was tested was an inactive mote which had stim electrode and ground testing leads connected to the corresponding testing leads of a functional mote that was acoustically driven and provided the stimulation current pulses. **g**, Impedance spectrum before and after PEDOT:PSS plating (device 1). (a-f): For each condition, the coloured line represents the trial-average voltage at each point in time, the width of the shaded error region is ± 1 s.d.. N, given on the left, is the number of stim events in each condition in each of the three panels to the right, respectively. A total of 26 pulses out of 1076 are not shown due to shorting region timing glitches. All replications for a given condition are separate stimulation events in a single *in vitro* experimental setup. N = 1 sweep for each condition in data in panel g.

StimDust adopts a design with monophasic stimulation and charge balance via electrode shorting (also known as passive recharge). This enables operation with a single power supply, saving area, power, and complexity on the volume- and power-constrained wireless device. This passive charge-balancing approach has been used successfully in clinical stimulators; Parastarfeizabadi and Kouzani 2017[6] write: "most of the available market-based open-loop and closed-loop DBS systems use a passive charge-balancing scheme." The passive charge balance leads to a discharge current as the remaining charge stored at the electrode interface flows through the tissue solution resistance during the shorting phase. Unlike many monophasic stimulators, StimDust has bipolar electrodes in close proximity and of identical area. As such, there is no asymmetry in current density at the electrodes, as there is for a monopolar stimulator which utilizes, for example, the large surface of an implanted titanium can as the return electrode. To ensure that this charge-balance current does not present any danger to the tissue, the tissue interface was modelled.

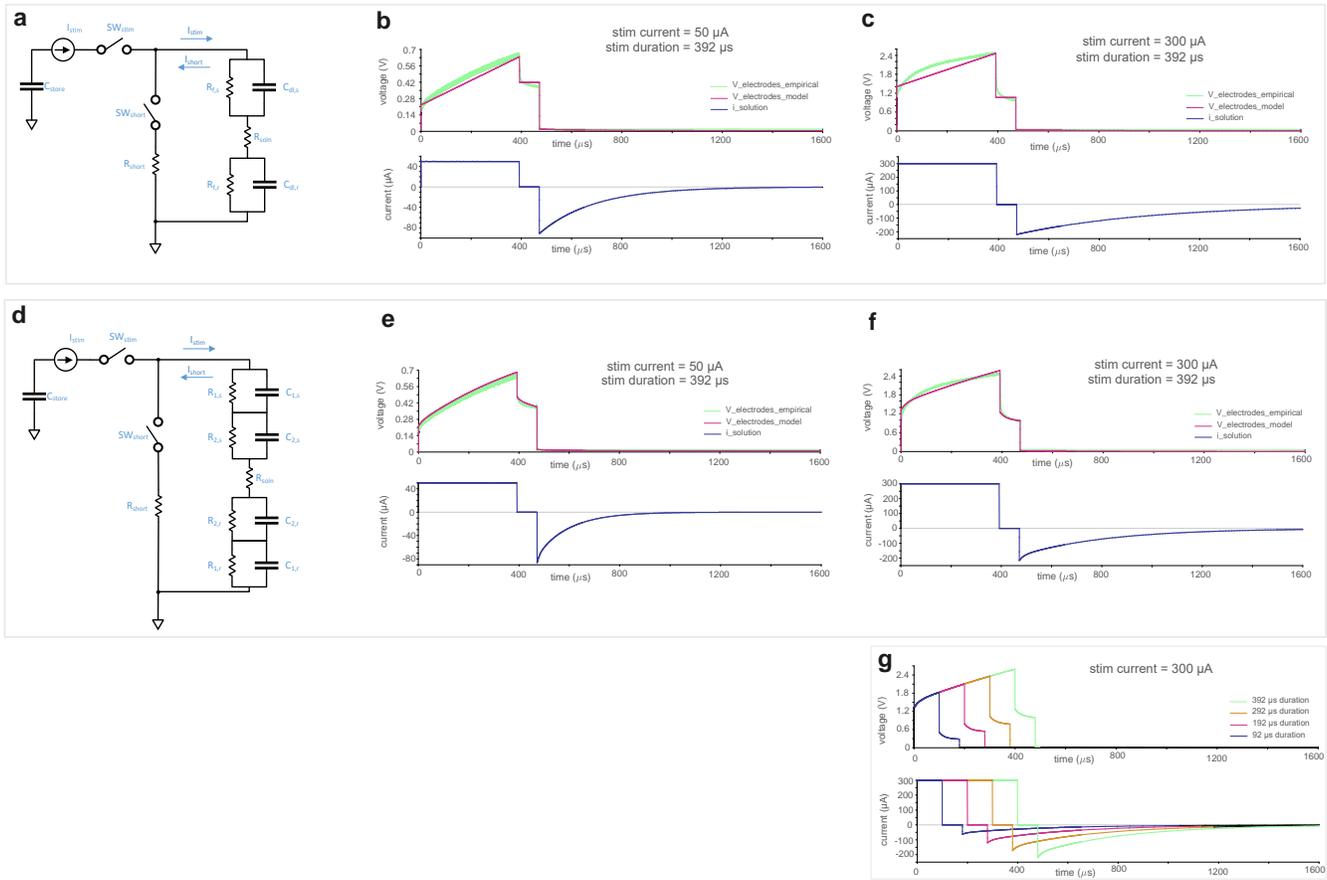

**Figure S9 | Monophasic stimulation and charge balancing**. **a**, simplified stimulator model (model A) with a simple electrode model: Cdl - Rs – Cdl. $R_{f,s}$ and $R_{f,r}$ are ignored in this model. **b**, 50 μA, 392 μs pulse fit to empirical data from animal D (parameters 95 nF, 4.4 kOhm). **c**, 300 μA, 392 μs pulse fit to empirical data from animal D (parameters 222 nF, 4.7 kΩ). **d**, simplified stimulator model (model B) with R1∥C1 - R2∥C2 - Rs - R2∥C2 - R1∥C1 electrode model. **e**, 50 μA, 392 μs pulse fit to empirical data from animal D (parameters 300 Ω, 40 nF, 8.6 kΩ, 65 nF,

4.2 kΩ). **f**, 300 µA, 392 µs pulse fit to empirical data from animal D (parameters 350 Ω, 55 nF, 6.0 kΩ, 190 nF, 4.4 kΩ). **g**, sweep of stimulation pulse duration from 100 µs to 600 µs (parameters: same as in f).

To model this, two circuits were simulated, each with a pulsed current source, an electrode-tissue equivalent circuit, and a shorting switch with shorting resistance (Fig S8). Stimulation current was applied for the (W) stim width phase; the electrodes were in a high-impedance state for the (G) interphase gap; and the shorting switch was closed for the (S) shorting phase. The resistance along the shorting path in the mote IC ($R_{short}$) is 130 Ω (the on-resistance of the shorting NMOS transistor (n-channel metal oxide semiconductor transistor) in Fig.3b). This reduces the maximum charge balance current and is independent of the tissue solution resistance. In model A, the electrode-tissue equivalent circuit is simplified to just a double-layer capacitance $C_{dl}$ at each electrode, and a solution resistance $R_{soln}$. This model is not perfect at fitting the shape of the stimulation pulse, but is widely used and easy to analyse. Model B considers a R1∥C1 - R2∥C2 at each electrode, with a solution resistance $R_{soln}$ in between. This equivalent circuit has been shown to fit the behaviour of neural stimulating electrodes[7]. It avoids an explicit constant phase element, as seen in some electrode impedance models which are analysed only in the frequency domain[8], because these elements take more parameters (RC elements) to model in the time domain[9].

Initial estimates of parameter values were generated using Zfit[10] on impedance spectroscopy data in PBS. Parameter estimates were then adjusted in a time-domain SPICE model to both peg the simulated electrode voltage at the end of the interphase gap to the empirical data from animal D and to fit the shape of the stimulation voltage as closely as possible. Through this fitting process, there was some variability or uncertainty in capacitor values, but little uncertainty in the series solution resistance. The maximum charge balance current depends almost entirely on the electrode voltage at the end of the interphase gap and the series solution resistance. The former is known empirically and the latter has a confident modelled value. Therefore, the estimates for maximum charge balance current are good. The estimates for the charge balance current time constant are slightly less certain due to lower confidence in the correct modelling of capacitive effects at the electrode interface.

| model | parameters | i_stim (μA) | Maximum charge balance current (μA) | Maximum charge balance current relative to stim current | Duration that charge balance current is greater than stim current (μs) | Time to discharge below 1 mV (μs) |
|---|---|---|---|---|---|---|
| A | 100 nF, 4.4kΩ | 50 | 83 | 167% | 116 | 1344 |
| A | 100 nF, 4.4kΩ | 100 | 131 | 131% | 61 | 1447 |
| A | 100 nF, 4.4kΩ | 150 | 162 | 108% | 18 | 1495 |
| A | 100 nF, 4.4kΩ | 200 | 162 | 81% | 0 | 1495 |
| A | 100 nF, 4.4kΩ | 250 | 185 | 74% | 0 | 1524 |
| A | 100 nF, 4.4kΩ | 300 | 211 | 70% | 0 | 1555 |
| A | 100 nF, 4.4kΩ | 350 | 228 | 65% | 0 | 1572 |
| A | 100 nF, 4.4kΩ | 400 | 245 | 61% | 0 | 1588 |
| A | 250 nF, 4.8 kΩ | 50 | 77 | 153% | 263 | 3657 |
| A | 250 nF, 4.8 kΩ | 100 | 131 | 131% | 167 | 3936 |
| A | 250 nF, 4.8 kΩ | 150 | 162 | 108% | 48 | 4066 |
| A | 250 nF, 4.8 kΩ | 200 | 162 | 81% | 0 | 4068 |
| A | 250 nF, 4.8 kΩ | 250 | 185 | 74% | 0 | 4147 |
| A | 250 nF, 4.8 kΩ | 300 | 211 | 70% | 0 | 4230 |
| A | 250 nF, 4.8 kΩ | 350 | 228 | 65% | 0 | 4278 |
| A | 250 nF, 4.8 kΩ | 400 | 245 | 61% | 0 | 4321 |
| B | 300 Ω, 40 nF, 8.6 kΩ, 65 nF, 4.2 kΩ | 50 | 89 | 179% | 52 | 729 |
| B | 300 Ω, 40 nF, 7.6 kΩ, 90 nF, 4.3 kΩ | 100 | 137 | 137% | 28 | 1068 |
| B | 300 Ω, 40 nF, 6.6 kΩ, 110 nF, 4.3 kΩ | 150 | 172 | 115% | 10 | 1304 |
| B | 325 Ω, 55 nF, 6.6 kΩ, 165 nF, 4.3 kΩ | 200 | 174 | 87% | 0 | 1969 |
| B | 330 Ω, 55 nF, 6.3 kΩ, 185 nF, 4.3 kΩ | 250 | 197 | 79% | 0 | 2230 |
| B | 350 Ω, 55 nF, 6.0 kΩ, 190 nF, 4.4 kΩ | 300 | 221 | 74% | 0 | 2352 |
| B | 350 Ω, 55 nF, 6.0 kΩ, 215 nF, 4.4 kΩ | 350 | 236 | 67% | 0 | 2676 |
| B | 350 Ω, 55 nF, 6.0 kΩ, 235 nF, 4.4 kΩ | 400 | 251 | 63% | 0 | 2957 |

**Table S2 | Charge balance current values from empirically pegged and fit simulations of the stimulation pulse.** The first set of simulations with model A use parameters from the range of fit parameters which yield maximal peak charge balance current. The second set of simulations with model A use parameters from the range of fit parameters which yield maximal discharge time.

Simulations were run and the charge balance current was measured, with key values extracted (Table S2). The simulations were performed with a stimulation pulse duration of 392 μs. This was the longest pulse duration utilized by StimDust, and is in the 'saturation' region of the recruitment curve by stimulation duration (Fig 8f). Peak charge balance current varies roughly linearly with stimulation pulse duration (Fig S8g), so pulses with duration less than 392 μs will have charge balance current values lower than those reported in Table S2.

Taken together, the data in Table S2 show that peak charge balance current never exceeds 251 μA, well below the nominal maximum stimulation current of 400 μA. Furthermore, the amount of time before charge balance current falls below the nominal stimulation current is substantially less than the stimulation pulse duration. All of these values would be less for shorter pulse durations. Importantly, most literature reporting the threshold for tissue damage is closely linked to charge per phase and charge density per phase[11]. The total charge transferred during the charge-balance phase is necessarily less than or equal to that during the stimulation phase. Stimulation protocols used in this work are plotted against the Shannon equation in Figure S9.

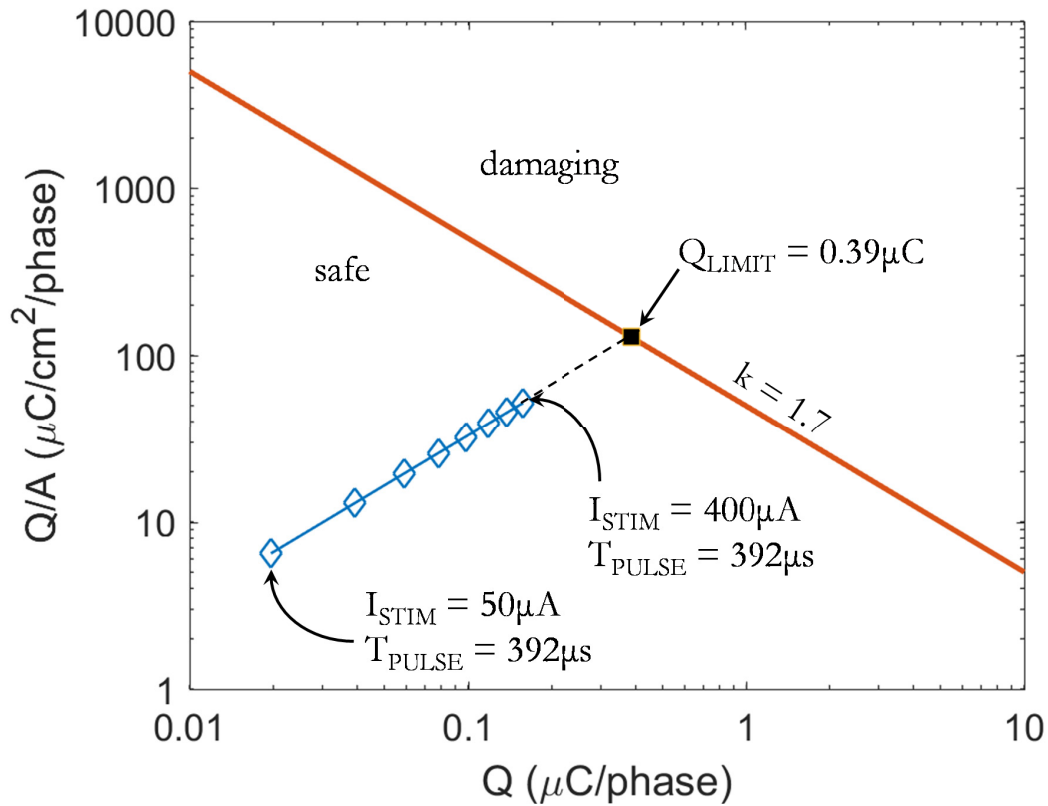

**Figure S10 | Charge (Q) vs. charge density (Q/A) for safe stimulation limits.** (Using k = 1.7). Data below the k = 1.7 line predict safe operation. The largest stimulation pulse used in this work is 40% of the limit. $T_{PULSE}$ is stimulation pulse duration.

The Shannon equation incorporates pulse width, current, and electrode area, but does not directly account for short, high intensity current pulses which can accompany passive recharge. Butterwick et. al. 2007 mapped the strength-duration dependence of damage thresholds for retinal cells. Assuming a damage threshold of 1 mA/cm², we estimated $R_{soln}$ for square electrodes *in vivo* and in saline for various areas to determine the charge balance current density ($J_{charge\_balance}$). We assumed a bipolar, symmetric electrode configuration and that the maximum stored electrode voltage was 1.2V (Figure S10).

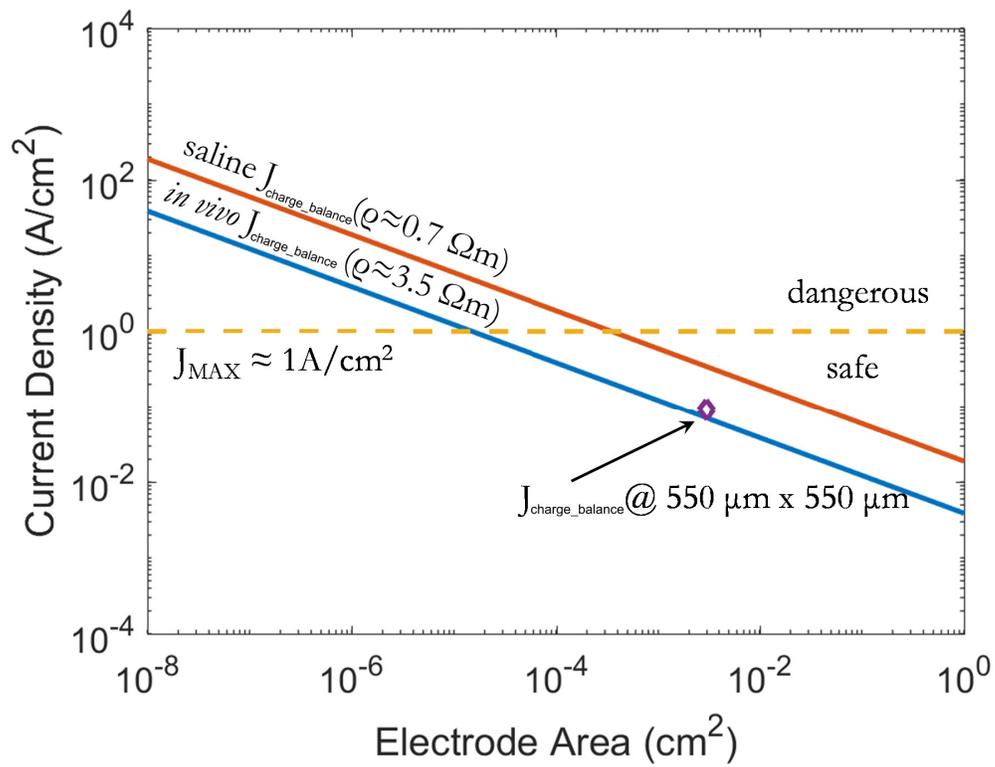

**Figure S11 | Charge balance current density vs. square electrode area in saline and *in vivo*.** As the stimulation and return electrodes scale to smaller, the discharge current can exceed the 1A/cm² damage threshold. The diamond represents the maximum discharge current for this work.

# Supplementary Section 5: Additional photographs

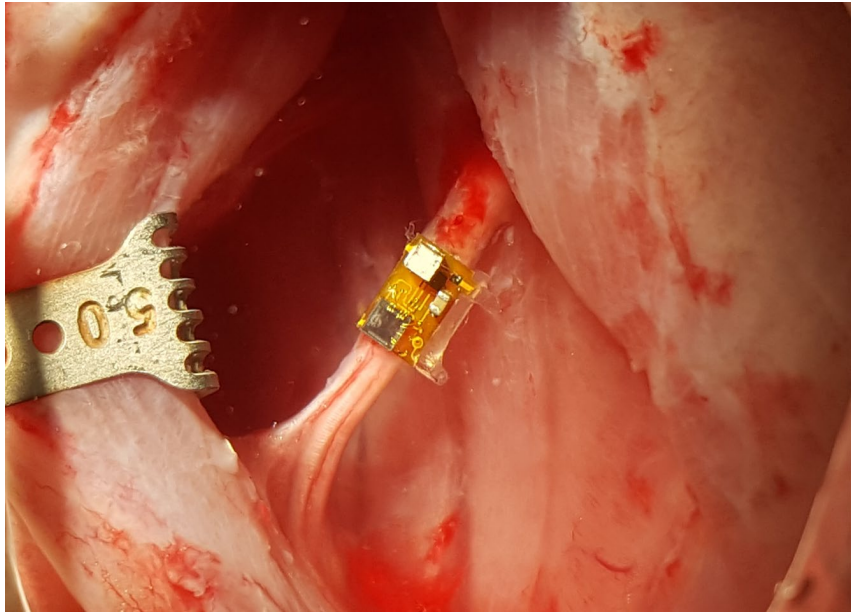

(a)

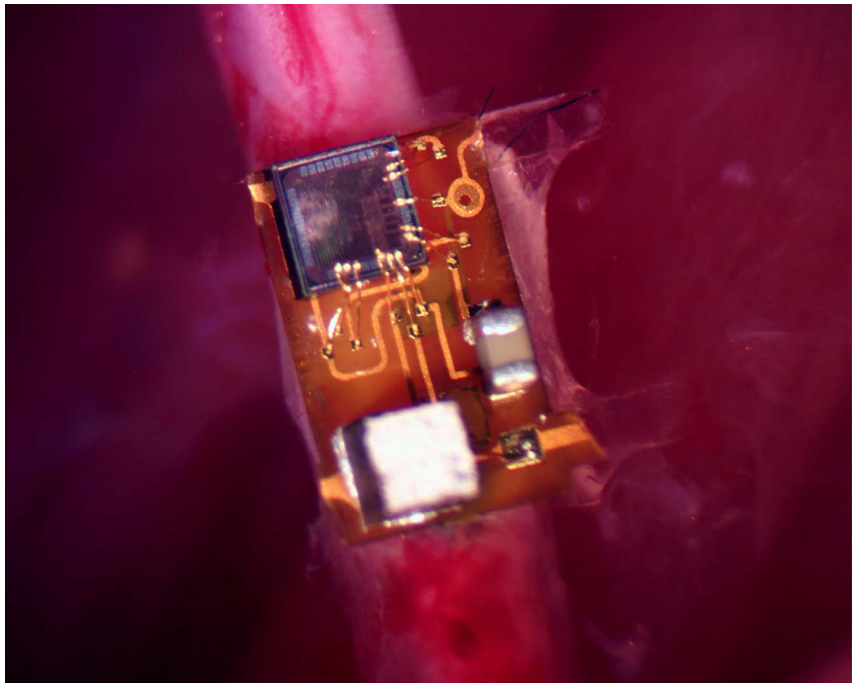

(b)

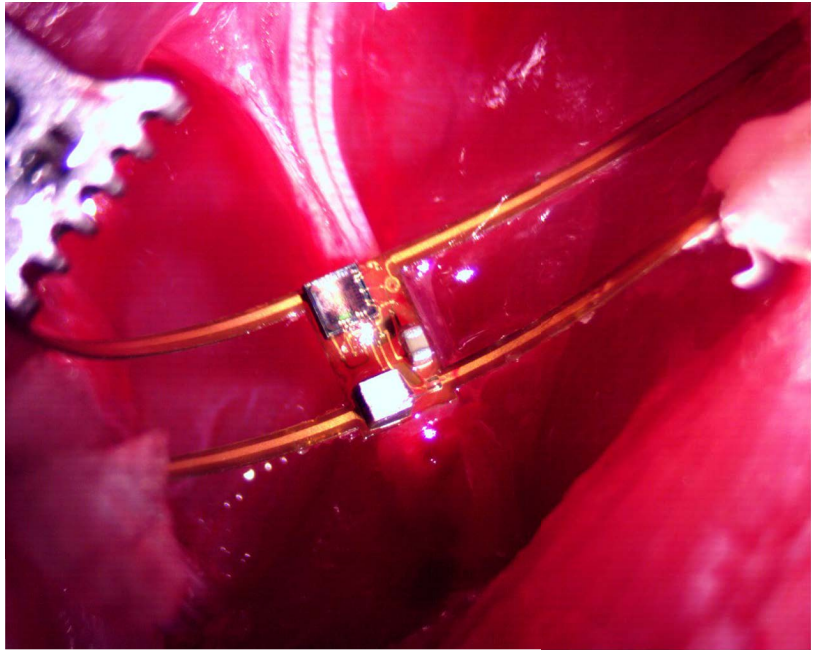

(c)

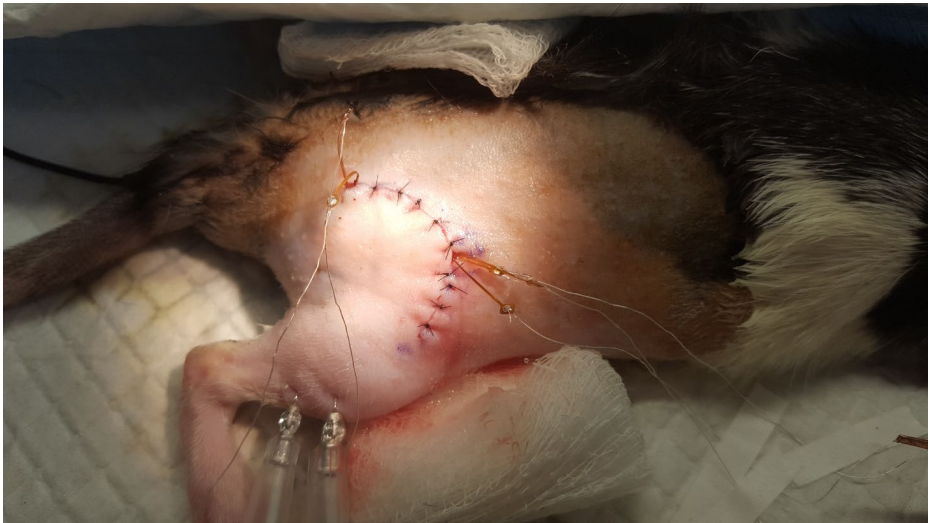

(d)

**Figure S12 | Alternate images for Fig 1b, Fig 1c, and Fig 1d.**
See main text Figure 1 caption for descriptions.

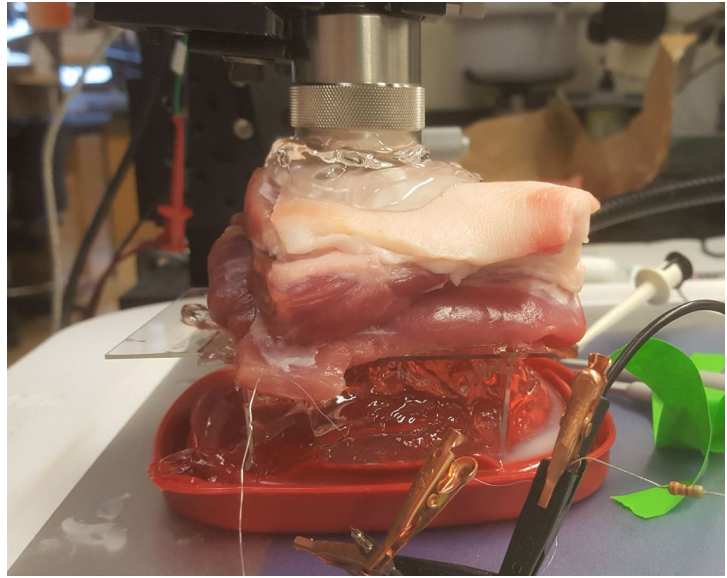

**Figure S13 | Alternate image for Fig 4c taken from a different angle**.

See main text Figure 4 for description.